\documentclass[journal]{IEEEtran}
\usepackage{amsmath,amsfonts,amssymb}
\usepackage{subcaption}
\usepackage{algorithmic}
\usepackage{algorithm}
\usepackage{array}
\usepackage{textcomp}
\usepackage{stfloats}
\usepackage{url}
\usepackage{verbatim}
\usepackage{graphicx}
\usepackage{cite}
\usepackage{hyperref}
\hypersetup{
	colorlinks=true,
	linkcolor=blue,
	citecolor=blue,
	urlcolor=black
}

\newtheorem{theorem}{Theorem}
\newtheorem{proposition}{Proposition}
\newtheorem{corollary}{Corollary}
\newtheorem{lemma}{Lemma}
\renewcommand{\IEEEQED}{\IEEEQEDopen}

\makeatletter
\let\HyThm@begintheorem\@begintheorem
\def\@begintheorem#1#2{%
	\phantomsection\HyThm@begintheorem{#1}{#2}}
\let\HyThm@opargbegintheorem\@opargbegintheorem
\def\@opargbegintheorem#1#2#3{%
	\phantomsection\HyThm@opargbegintheorem{#1}{#2}{#3}}
\makeatother

\DeclareMathOperator{\Ex}{\mathbb{E}}
\DeclareMathOperator{\Pb}{\mathbb{P}}

\DeclareMathOperator{\EX}{\mathbb{E}}
\DeclareMathOperator{\PR}{\mathbb{P}}
\newcommand{\mc}[1]{\mathcal{#1}}
\newcommand{\bs}[1]{\boldsymbol{#1}}

\newcommand{\algoline}[2]{\STATE {#1}:\hspace{4pt}#2}

\begin{document}

\title{Drift Plus Optimistic Penalty - A Learning Framework for Stochastic Network Optimization}

\author{
	\IEEEauthorblockN{
		Sathwik Chadaga,~\IEEEmembership{Student Member,~IEEE,} and Eytan Modiano,~\IEEEmembership{Fellow,~IEEE.}}\\
	\IEEEauthorblockA{Laboratory for Information and Decision Systems, Massachusetts Institute of Technology, Cambridge, MA}
	\vspace{-16pt}
\thanks{This work was supported by NSF grants CNS-2148183, CNS-2106268, and CNS-2148128.}
}

\markboth{Accepted for publication in IEEE/ACM Transactions on Networking}%
{Chadaga \MakeLowercase{\textit{et al.}}: Drift Plus Optimistic Penalty - A Learning Framework for Stochastic Network Optimization}

\IEEEpubid{0000--0000/00\$00.00~\copyright~2021 IEEE}

\maketitle

\begin{abstract}
	We consider the problem of joint routing and scheduling in queueing networks, where the edge transmission costs are unknown. At each time-slot, the network controller receives noisy observations of transmission costs only for those edges it selects for transmission. The network controller's objective is to make routing and scheduling decisions so that the total expected cost is minimized. This problem exhibits an exploration-exploitation trade-off, however, previous bandit-style solutions cannot be directly applied to this problem due to the queueing dynamics. In order to ensure network stability, the network controller needs to optimize throughput and cost simultaneously. We show that the best achievable cost is lower bounded by the solution to a static optimization problem, and develop a network control policy using techniques from Lyapunov drift-plus-penalty optimization and multi-arm bandits. We show that the policy achieves a sub-linear regret of order $O(\sqrt{T}\log T)$, as compared to the best policy that has complete knowledge of arrivals and costs. Finally, we evaluate the proposed policy using simulations and show that its regret is indeed sub-linear.
\end{abstract}


\section{Introduction}\label{sec:introduction}
\IEEEPARstart{}Stochastic network optimization refers to the problem of making routing and scheduling decisions in network systems in order to optimize objectives such as cost, utility, and throughput. It is fundamental in designing good network resource allocation strategies and has applications in many fields such as communications \cite{xiao_known_costs, cruz_known_costs},  cloud computing \cite{data_centers, data_centers_queues}, and content delivery \cite{content_distribution_queues, streaming_ghaderi}.
Early studies on network optimization \cite{xiao_known_costs, cruz_known_costs, data_centers} consider a static version of this problem, where traffic is modeled as static flows.
However, practical network systems are not static and require queue management due to the stochastic nature of traffic. Therefore, their control involves maintaining low queueing delays to ensure high throughput. 

In \cite{uysal_single_server,fu_single_server}, throughput optimal policies are designed for power constrained wireless systems with known arrival rates.
When arrival statistics are unknown, the Lyapunov optimization technique \cite{tassiulass_bp, neely_modiano_2005} is commonly used to develop throughput-optimal policies.
Lyapunov techniques are also used to design congestion control and scheduling algorithms in network utility maximization problems \cite{lin_shroff_num, srikant_atilla_num}.
A policy that jointly optimizes energy and throughput is proposed in \cite{neely_known_costs}. It uses the Lyapunov drift-plus-penalty minimization technique to demonstrate a trade-off between throughput and energy consumption.
This technique has since been used to design throughput-optimal and minimum-cost policies in many other fields. For example, Lyapunov optimization has been used to develop cost effective battery management schemes in data centers \cite{data_centers_queues} and content distribution strategies in cloud infrastructures \cite{content_distribution_queues}. A similar power control strategy for wireless networks with batteries is studied in \cite{tassiulas_power_constraint}. Finally, \cite{hop_minimization} combines Lyapunov minimization and shortest-path routing to design a minimum hop routing scheme. All of the above works assume that the transmission costs are known in advance.

In applications where costs represent quantities such as power consumption \cite{unknown_costs_eg_energy, battery_management_learning}, monetary cost \cite{unknown_costs_eg_cost, battery_management_learning},  server load \cite{unknown_cost_eg_load}, or quality of service \cite{unknown_cost_eg_qos}, costs may initially be unknown and need to be learned through feedback. Moreover, the costs are revealed only partially depending on the choice of action.
This is similar to the multi-arm bandit problem \cite{lai_bandits}, where a controller has to pull the best arm among multiple arms, each associated with an unknown cost. At each time, the controller receives noisy cost observation for only the pulled arm. Hence, it faces a trade-off between pulling the best arm according to past observations, versus pulling an under-explored arm in the hope of finding better arms. This is called the exploration-exploitation trade-off.
A widely used strategy to deal with this trade-off is the upper-confidence-bound strategy \cite{auer_bandits, bianchi_bandits} where, arms are chosen based on optimistically biased cost estimates, which allow exploration by lowering the costs of under-explored arms.

Routing in networks with unknown costs can also be formulated as a multi-arm bandit problem with each route acting as an arm. However, this method is not scalable as the number of routes (arms) grows exponentially with network size. Alternatively, one can exploit the fact that paths share edges in a network and hence path costs are dependent. Consequently, an efficient exploration basis called barycentric spanner that can be computed in polynomial time is proposed in \cite{kleinberg_barycentric}. Using this, the confidence-bound style exploration is extended to shortest path routing in \cite{dani_linearbandits, tsitsiklis_bandits, abbasi_bandits}. 
\IEEEpubidadjcol

These bandit-type solutions cannot be extended to queueing networks directly. In the traditional bandits problem, the best arm with the least underlying cost is optimal. The solution involves exploring enough arms and converging onto the best arm quickly. However, in queueing networks, the arrival rate is often larger than the capacity of a single path. Hence, minimizing the cost alone will cause queue backlogs to build up along low cost paths, which could lead to instability.
Thus, the network controller must optimize both cost and throughput simultaneously.

The problem of optimal routing and scheduling in a single-hop queueing system with unknown costs is considered in \cite{xinzhe_parallel} where, a priority scheme based on queue lengths and upper-confidence-bound cost estimates is designed. Despite having an instance-dependent logarithmic regret, this policy is limited to single-hop systems.
Further, similar constrained bandit problems are studied in \cite{bandits_fairness_contraints, bandits_with_general_constraints} using Lyapunov drift analysis, where constraint  violations are tracked using virtual queues. Extending these policies to tackle the optimal routing problem requires bounding physical queues rather than virtual queues. Additionally, similar to \cite{xinzhe_parallel},  these policies are limited to single-hop (virtual) queues. In multi-hop networks, the packets routed through one path will affect queue backlogs of other overlapping paths. Hence, extending these works to multi-hop networks is non-trivial. Finally, \cite{amer_online_routing} addresses multi-hop routing through traffic splitting at the source, by using an optimistically estimated version of a traffic splitting metric from  \cite{hop_minimization}. However, the regret is defined as a function of only this traffic splitting metric and not as a function of the total cost. Hence, this does not address the total cost minimization problem.

In this paper, we consider the problem of optimal routing and scheduling in multi-hop stochastic queueing networks. We assume that the arrival rates and edge transmission costs are unknown. Instead, the control policy has to make transmission decisions based only on the queue backlogs and past cost observations. We seek to design a policy with sub-linear regret, where the regret is defined as the gap between the policy's expected cost and the cost of an optimal policy with complete knowledge of arrivals and costs. In a preliminary version of this work \cite{my_infocom2025_paper}, we considered the single-commodity version of this problem, where we proposed a policy that has $O(T^{2/3})$ regret, where $T$ is the time horizon. In this paper, we extend this policy to multi-commodity networks, and we also improve the policy's regret performance to $O(\sqrt{T}\log T)$. We summarize our contributions below.
\begin{itemize}
	\item We define a novel cost metric in terms of both transmission costs and queue backlogs that allows us to jointly optimize cost and throughput. We show that the best achievable cost of any policy is lower bounded by the solution to a static optimization problem. 
	\item We develop a network control policy by combining ideas from Lyapunov drift-plus-penalty minimization technique and the upper-confidence-bound algorithm. Further, we show that the proposed policy has a sub-linear regret of order $O(\sqrt{T}\log T)$, where $T$ is the time horizon.
	\item We evaluate the proposed policy using simulations and show that it indeed has a sub-linear regret. We also simulate an oracle policy that knows the underlying costs exactly and show that the proposed policy's backlog and cost performance approaches the oracle's performance.
\end{itemize}

The rest of the paper is organized as follows. We describe our model and formulate the network control problem in Section \ref{sec:multi_user}. We derive a lower bound on the best achievable cost in Section \ref{sec:static_lower_bound}.  We discuss the proposed Drift Plus Optimistic Penalty policy in Section \ref{sec:algorithm} and analyze the policy's regret in Section \ref{sec:regret_analysis}. Finally, we present the simulation results in Section \ref{sec:results} and conclusions in Section \ref{sec:conclusions}.

\section{Problem Formulation}\label{sec:multi_user}

\paragraph{Network and Traffic Model} We consider a multi-hop network $G = (\mc{N},\mc{E})$, where $\mc{N}$ is the set of nodes and $\mc{E}$ is the set of directed edges.
The network operates at discrete time-slots $t=1,2,...,T$, where $T$ is the given\emph{ time horizon}.
At each time $t$ and $\forall k\in\mc{N},\;\forall i\in \mc{N}$, there are $a_{ik}(t)$ packets that arrive at node $i$ destined to node $k$. We refer to the packets that are destined to node $k\in\mc{N}$ as \textit{commodity-$k$ packets}.
The arrivals $a_{ik}(t)$'s are independent bounded random variables with unknown arrival rates $\lambda_{ik}:=\Ex[a_{ik}(t)]$ packets/slot. We denote the set of all arrival rates by $\bs{\lambda}$.
The network maintains first-in-first-out queues to buffer incoming packets at every node. We denote by $Q_{ik}^\pi(t)$ the \emph{queue backlog} of commodity $k\in\mc{N}$ at node $i\in \mc{N}$, and by $\bs{Q}^\pi(t)$ the set of all queue backlogs at time $t$. We use bold notations to denote sets and vectors throughout the rest of the paper.
A given control policy $\pi$ observes the queue backlogs at each time-slot and decides the number of packets of each commodity to be transmitted on each edge. We denote by $\mu_{ijk}^\pi(t)$ policy $\pi$'s \emph{planned transmission} in packets/slot on edge $(i,j)\in \mc{E}$ for commodity $k\in\mc{N}$, and by $\bs{\mu}^\pi$ the set of all planned transmissions at time $t$. We denote the set of outgoing neighbors of any node $i\in \mc{N}$ by $\mc{N}_i := \{j\in \mc{N} : (i,j)\in \mc{E}\}$. The queue backlog evolution  can now be expressed as follows, $\forall k\in\mc{N},\;\forall i\in \mc{N}$ such that $i\neq k$, and $\forall t=1,2,...,T$,
\begin{multline}\label{eq:multi_queue_evol_i}
	Q^\pi_{ik}(t+1) \leq \left[Q^\pi_{ik}(t) - \sum_{j\in \mc{N}_i}\mu^\pi_{ijk}(t) \right]^+ \\
	+ \sum_{j:i\in \mc{N}_j}\mu^\pi_{jik}(t) + a_{ik}(t)
\end{multline}
where, $[\cdot]^+ = \max\{\cdot,0\}$. Finally, we assume that packets exit the network immediately when they reach their destination nodes. Hence $\forall t=1,...,T$, $\forall k\in \mc{N}$, we have $Q^\pi_{kk}(t) = 0$.

Notice that the queue evolution expression (\ref{eq:multi_queue_evol_i}) is an inequality. This is because, for each incoming edge $(j,i) \in \mc{E}$ and for each commodity $k\in\mc{N}$ at node $i$, the queue $Q_{jk}^\pi(t)$ may not have enough buffered packets to support the planned transmission $\mu_{jik}^\pi(t)$. As a result, the actual number of packets transmitted on some edges may be less than the planned number of transmissions. We denote by $\tilde{\mu}^\pi_{ijk}(t)$ the \emph{actual transmission} in packets/slot of commodity $k\in\mc{N}$ on edge $(i,j)\in \mc{E}$ at time $t$, and by $\bs{\tilde{\mu}}^\pi(t)$ the set of all actual transmissions at time $t$. Note that the actual transmissions are constrained by $\sum_{j\in \mc{N}_i}\tilde{\mu}^\pi_{ijk}(t) \leq Q_{ik}^\pi(t), \; \forall i\in \mc{N}, \forall k\in\mc{N}$ and $ \tilde{{\mu}}^\pi_{ijk}(t) \leq {\mu}^\pi_{ijk}(t), \; \forall (i,j)\in \mc{E}, \forall k \in \mc{N}$.

\paragraph{Stability Region} We aim to keep the queue backlogs bounded, so that our throughput is equal to the arrival rate, using the notion of \emph{rate stability} from \cite{neely_book}. A queueing network with backlogs $\bs{Q}^\pi(t)$ is said to be rate stable under policy $\pi$ if the queue backlogs satisfy $$\lim_{T\rightarrow\infty} \frac{1}{T} \sum_{i\in \mc{N}} \sum_{k\in \mc{N}}  \Ex[Q^\pi_{ik}(T)] = 0.$$

Further, each edge $(i,j)\in \mc{E}$ has a finite capacity denoted by $\mu^{max}_{ij}$. A policy $\pi$ is feasible if its transmission decisions satisfy $\forall t, \forall(i,j)\in \mc{E}, \; 0 \leq \sum_{k\in \mc{N}} \mu^\pi_{ijk}(t) \leq \mu^{max}_{ij}$. Let $\Pi^*$ be the collection of all feasible control policies, including the policies with knowledge of future arrivals.
We define the \emph{stability region} $\Lambda(G)$ as the collection of arrival rates $\bs{\lambda}$ for which there exists a policy $\pi\in\Pi^*$ that keeps the system stable. In \cite{neely_first_book}, it was shown that  the stability region $\Lambda(G)$ can be characterized as the collection of all arrival rates $\bs{\lambda}$ for which there exists feasible flows $\bs{\mu} := \{\mu_{ijk} : {(i,j)\in \mc{E}, k\in\mc{N}}\}$ that satisfy the conditions
\begin{equation}
	\begin{Bmatrix}
		\sum_{j:i\in \mc{N}_j} \mu_{jik} + \lambda_{ik} \leq \sum_{j\in \mc{N}_i} \mu_{ijk}, \; \forall i\neq k, \\ 
		\sum_{k \in \mc{N}} \mu_{ijk} \leq \mu_{ij}^{max}, \; \forall (i,j)\in \mc{E}, \\
		\mu_{ijk} \geq 0, \; \forall (i,j)\in \mc{E}, k\in\mc{N}.
	\end{Bmatrix} \label{eq:capacity_region_multi}
\end{equation}

Formally, the stability region $\Lambda(G)$ is defined as the collection of all arrival rate vectors $\bs{\lambda}$ such that $\lambda_{ik} \geq 0, \; \forall i,k\in\mc{N}$ and  $\exists \; \bs{\mu}$ satisfying (\ref{eq:capacity_region_multi}),
$
	\Lambda(G) := \left\{\bs{\lambda} \geq 0: \exists \; \bs{\mu} \text{ satisfying (\ref{eq:capacity_region_multi})}\right\}.
$
Throughout this paper, we assume that our arrival rates are strictly within the stability region $\bs{\lambda} \subset \Lambda(G)$. Formally, we assume $\exists \, \epsilon>0$ and $\exists \, \{\mu_{ijk} : {(i,j)\in \mc{E}, k\in\mc{N}}\}$ such that,
\begin{equation}
	\begin{Bmatrix}
		\sum_{j:i\in \mc{N}_j} \mu_{jik} + \lambda_{ik} +\epsilon \leq \sum_{j\in \mc{N}_i} \mu_{ijk}, \; \forall i\neq k, \\ 
		\sum_{k \in \mc{N}} \mu_{ijk} \leq \mu_{ij}^{max}, \; \forall (i,j)\in \mc{E}, \\
		\mu_{ijk} \geq 0, \; \forall (i,j)\in \mc{E}, k\in\mc{N}.
	\end{Bmatrix} \label{eq:multi_capacity_interior}
\end{equation}




	\paragraph{Cost Structure} Each edge $(i,j)$ has an unknown \emph{transmission cost} of $c_{ij}$ per packet. Hence, a given policy $\pi$ incurs a total transmission cost of $ \sum_{t=1}^T \sum_{(i,j)\in \mc{E}} \sum_{k\in \mc{N}} \tilde{\mu}^\pi_{ijk}(t)c_{ij} $. We assume that these costs are bounded by $0 \leq c_{ij} \leq C_{max}$. Further, minimizing the transmission cost alone may cause network instability. For example, a bad policy can attain zero cost by simply not transmitting any packets and letting the queues build up. To avoid this, we include the following queue backlog penalty. The policy incurs a \emph{terminal backlog cost} $C_{B} \geq 0$ for each undelivered packet at the end of the time horizon $T$. Hence, the policy incurs a total backlog cost of $C_{B}\sum_{i\in \mc{N}} \sum_{k\in\mc{N}} Q_{ik}^\pi(T)$. Intuitively, we should pick a large terminal cost $C_B$ so that the policy is encouraged to deliver packets to the destination. Otherwise, the policy can let queues build up and cause instability. This terminal cost can also be interpreted as the cost incurred to deliver the remaining packets at the end of time horizon $T$ to the destination. We will discuss a good choice of $C_B$ in Section \ref{sec:static_lower_bound}. In summary, the total cost $C^\pi(T)$ incurred by a given control policy $\pi$ is defined as follows:
	\begin{equation}\label{eq:multi_actual_costs}
		C^\pi(T) := \sum_{t=1}^T \sum_{(i,j)\in \mc{E}} \sum_{k\in\mc{N}} \tilde{\mu}^\pi_{ijk}(t)c_{ij} + C_{B}\sum_{i\in \mc{N}} \sum_{k\in\mc{N}} Q_{ik}^\pi(T).
	\end{equation}
	As actual transmissions $\bs{\tilde{\mu}}$ are constrained both by edge capacities and queue states, working with them complicates the analysis. Hence, we simplify our analysis by expressing the total cost in terms of the planned transmissions $\bs{\mu}$ instead. By definition, we have  $\bs{\tilde{\mu}} \leq \bs{\mu}$. And since the costs are non-negative, we have
	\begin{equation}\label{eq:multi_planned_costs}
		C^\pi(T) \leq \sum_{t=1}^T \sum_{(i,j)\in \mc{E}} \sum_{k\in\mc{N}} {\mu}^\pi_{ijk}(t)c_{ij} + C_{B}\sum_{i\in \mc{N}} \sum_{k\in\mc{N}} Q_{ik}^\pi(T).
	\end{equation}
	This simplifies our analysis as the planned transmissions are only constrained by edge capacities and not by the queue sizes. However, note that the exact cost expression (\ref{eq:multi_actual_costs}) will still be required to prove Theorem \ref{thm:static_lower_bound} in Section \ref{sec:static_lower_bound}.

	As the costs $c_{ij}$'s are initially unknown, the policy relies on the feedback it receives depending on its actions. At each $t$, the policy receives a \emph{noisy observation} $ \widetilde{c}_{ij} (t)$ for each edge on which it transmitted any packets. Formally, let $e_{ij}(t) \in \{0,1\}$ be an indicator variable that tracks whether an observation was received for edge $(i,j)$ at time $t$. For each edge $(i,j)\in \mc{E}$,
	\begin{equation*}
		e_{ij}(t) = \begin{cases}
			1 & \text{if } \exists \; k\in\mc{N} : \; \mu^\pi_{ijk}(t) >0, \\
			0 &  \text{otherwise.}
		\end{cases}
	\end{equation*}
	The received partial noisy observations $ \widetilde{c}_{ij} (t)$'s at time $t$ are
	\begin{equation*}
		 \widetilde{c}_{ij} (t) := c_{ij} + \eta_{ij}(t), \; \forall (i,j)\in \mc{E} : e_{ij}(t) = 1
	\end{equation*}
	where, $\eta_{ij}(t)$'s are zero mean $\sigma$-sub-Gaussian\footnote{A random variable $X$ is $\sigma$-sub-Gaussian if $\Ex[e^{\zeta(X-\Ex[X])}]\leq e^{\frac{\sigma^2\zeta^2}{2}}, \forall \zeta$.} random variables that are independent across time slots. As discussed before, the neighboring queues of some edges with $\mu^\pi_{ijk}(t) >0$ may not have enough packets to support the planned transmissions. On these edges, we assume that the policy can send dummy packets, no more than $\mu^\pi_{ijk}(t)$, and still observe the costs. Note that these dummy packets do not contribute to the queue evolution but they do incur transmission costs, which is captured in (\ref{eq:multi_planned_costs}). Also note that this assumption does not affect our analysis as we have expressed the cost as a function of planned transmissions in (\ref{eq:multi_planned_costs}) rather than the actual transmissions.
	Denote by $\Pi^*$ the collection of all policies, including those with knowledge of costs and future arrivals. We define the \emph{regret} of policy $\pi$ as
	\begin{equation*}
		R^\pi(T) := \Ex [C^\pi(T)] - \inf_{\pi^*\in\Pi^*} \Ex[ C^{\pi^*}(T)].
	\end{equation*}
	where, the expectation is taken over the randomness in arrivals and possibly in policy's actions. Note that, unlike the best policy $\pi^*$, the policies we consider will not have access to the cost values, future arrivals, and even the arrival rates. Let $\Pi$ be the collection of admissible policies that do not know the costs $c_{ij}$'s, do not know the arrival rate $\lambda$, and make causal control decisions. We now state our objective as follows.
	\paragraph*{Objective} Find a policy $\pi\in\Pi$ that has sub-linear regret
	\begin{equation*}
		\lim_{T\rightarrow\infty}\frac{R^\pi(T)}{T} = 0.
	\end{equation*}

\section{Static Lower Bound on the Optimal Cost}\label{sec:static_lower_bound}
In this section, we obtain a lower bound on the optimal cost, using a static flow version of the problem. This bound will be useful for the regret analysis of our proposed policy. Consider the optimization problem $\mc{P}$.
\begin{gather*}
	\mc{P}(\bs{\lambda}) := \; \min_\mu \sum_{(i,j)\in \mc{E}} \sum_{k\in\mc{N}} \mu_{ijk}c_{ij}, \\
		\text{subject to }\sum_{j:i\in \mc{N}_j} \mu_{jik} + \lambda_{ik} \leq \sum_{j\in \mc{N}_i} \mu_{ijk}, \; \forall i\neq k, \\ 
		\sum_{k \in \mc{N}} \mu_{ijk} \leq \mu_{ij}^{max}, \quad \forall (i,j)\in \mc{E}. \\
		\mu_{ijk} \geq 0, \quad \forall (i,j)\in \mc{E}, \; \forall k\in\mc{N}.
\end{gather*}

In the following theorem, we show that for a large enough terminal backlog cost $C_B$, the optimal value of $\mc{P}$ lower-bounds the best policy's cost. Since $\mc{P}$ is bounded, it has at least one optimal solution. Let $\bs{\mu}^{stat} := \{\mu_{ijk}^{stat} :  {(i,j)\in \mc{E}, k\in\mc{N}}\}$ be an optimal solution to $\mc{P}$.
\begin{theorem}\label{thm:static_lower_bound} \emph{(Static Lower Bound)} There exists a finite constant $C_L$ that is only a function of the network topology and transmission costs, such that, for $C_B \geq C_L$, we have
	\begin{equation*}
		\inf_{\pi^*\in\Pi^*} \Ex[C^{\pi^*}(T)] \geq T \sum_{(i,j)\in \mc{E}} \sum_{k\in\mc{N}} \mu^{stat}_{ijk}c_{ij}.
	\end{equation*}
\end{theorem}




\emph{Proof:}
We first define some variables that will be useful for the proof. Recall that the number of commodity $k$ packets transmitted on any edge $(i,j)$ under policy ${\pi^*}$ at time $t$ is $\tilde{\mu}^{{\pi^*}}_{ijk}(t)$. Among these packets, let $\hat{\mu}^{{\pi^*}}_{ijk}(t)$ be the number of packets that got delivered to the destination by the end of time horizon $T$. Similarly, among the $a_{ik}(t)$ packets that arrived at time $t$, let $\hat{a}^{\pi^*}_{ik}(t)$ be the packets that reached their destination by $T$ under policy $\pi^*$. We define the average effective rate as $\bar{\mu}^{\pi^*}_{ijk} := \EX \big[\sum_{t=1}^T \hat{\mu}^{\pi^*}_{ijk}(t)/T \big]$, and define the average effective arrivals as $\bar{\lambda}_{ik}^{\pi^*} := \EX \big[\sum_{t=1}^T \hat{a}^{\pi^*}_{ik}(t)/T \big]$.
We show certain properties of these quantities in the lemma below.

\begin{lemma}\label{lemma:mean_rates_property} The average effective rates $\bar{\mu}_{ijk}^{\pi^*}$ and the average final backlogs $Q_{ik}^{\pi^*}(T)$ satisfy
	\begin{gather}
		\sum_{j:i\in \mc{N}_j} \bar{\mu}^{\pi^*}_{jik} + \bar{\lambda}^{\pi^*}_{ik} = \sum_{j\in \mc{N}_i} \bar{\mu}^{\pi^*}_{ijk}, \; \forall i\neq k. \label{eq:mean_rates} \\
		T\sum_{i\in \mathcal{N}}\sum_{k\in \mathcal{N}} \big( \lambda_{ik} - \bar{\lambda}^{\pi^*}_{ik} \big ) = \sum_{i\in \mathcal{N}} \sum_{k\in \mathcal{N}} \Ex[Q_{ik}^{\pi^*}(T)] \label{eq:mean_rates_arrivals}
	\end{gather}
\end{lemma}

We prove this lemma in Appendix \ref{proof:mean_rates_property}. We now express the best policy's cost in terms of $\bar{\mu}_{ijk}^{\pi^*}$ and $Q_{ik}^{\pi^*}(T)$. We know that not all of the $\tilde{\mu}^{{\pi^*}}_{ijk}(t)$ packets transmitted at time $t$ reached their destinations by the end of time horizon $T$. Hence, the number of delivered packets is lower than the number of transmitted packets i.e. the packets counted by $\hat{\mu}^{{\pi^*}}_{ijk}(t)$ is a subset of the packets counted by $\tilde{\mu}^{{\pi^*}}_{ijk}(t)$, and we have $\tilde{\mu}^{{\pi^*}}_{ijk}(t) \geq \hat{\mu}^{{\pi^*}}_{ijk}(t)$. Using this in the cost definition (\ref{eq:multi_actual_costs}),
\begin{align}
	\Ex[C^{\pi^*}(T)] \geq T \sum_{(i,j)\in \mc{E}} &\sum_{k\in\mc{N}} \bar{\mu}^{\pi^*}_{ijk}c_{ij} \nonumber \\
		 &+ C_{B}\sum_{i\in \mc{N}} \sum_{k\in\mc{N}} \Ex[Q_{ik}^{\pi^*}(T)]. \label{eq:cost_inequality_1}
\end{align}
%

The non-zero number of undelivered packets left at the end of time horizon $\sum_{i\in \mc{N}} \sum_{k\in\mc{N}} \Ex[Q_{ik}^{\pi^*}(T)]$ implies that the effective rates $\{\bar{\mu}^{\pi^*}_{ijk}\}$ did not support the arrival rate $\bs{\lambda}$ entirely. In other words, the rates $\{\bar{\mu}^{\pi^*}_{ijk}\}$ may not lie in the feasibility region of $\mathcal{P}({\bs{\lambda}})$. However, from the equality in (\ref{eq:mean_rates}), we can see that the rates  $\{\bar{\mu}^{\pi^*}_{ijk}\}$ do indeed support the effective arrival rate $\bar{\bs{\lambda}}$. Formally, the rates $\{\bar{\mu}^{\pi^*}_{ijk}\}$ lie within the feasibility region of $\mathcal{P}(\bar{\bs{\lambda}})$. Hence, we have $\sum_{ij} \sum_{k} \bar{\mu}^{\pi^*}_{ijk} c_{ij} \geq \mc{P}(\bar{\bs{\lambda}})$. Plugging this in (\ref{eq:cost_inequality_1}), using (\ref{eq:mean_rates_arrivals}), and subtracting $ T\mc{P}(\bs{\lambda}) $ from both sides, 
\begin{multline}\label{eq:cost_inequlality}
\Ex[C^{\pi^*}(T)] - T \mc{P}(\bs{\lambda}) \\
\geq T \left(\mc{P}(\bar{\bs{\lambda}}) -\mc{P}(\bs{\lambda}) \right)+ TC_{B}\sum_{i\in \mc{N}} \sum_{k\in\mc{N}}  \big( \lambda_{ik} - \bar{\lambda}_{ik}^{\pi^*} \big).
\end{multline}

To complete the proof, we need to prove that there exists a finite $C_L$ such that the right hand side of the above inequality is non-negative when $C_B\geq C_L$. We first prove this for the special case of single commodity networks with one source-destination pair, where an intuitive result can be obtained. In this special case, we will see that the result can be achieved by setting $C_L$ to the cost of most expensive path in the residual graph. We defer the proof of general multi-commodity case to Appendix \ref{proof:static_lower_bound_multi}. 

\emph{Single-Commodity Special Case:}  Assume that the network has a single source-destination pair $s$-$d$. Packets of a single commodity arrive at node $s\in\mc{N}$ destined to node $d\in\mc{N}$ at each time-slot at rate $\Ex[a_{sd}(t)] = \lambda$, and $a_{ik}(t) = 0$ for $i\neq s$. Define the single-commodity version of effective arrival rate as $\bar{\lambda} := \Ex[\hat{a}^{\pi^*}_{sd}(t)]$. In this special case, the vectors $\bs{\lambda}$ and $\bar{\bs{\lambda}}$ can be replaced by scalars $\lambda$ and $\bar{\lambda}$ respectively. Hence, the inequality (\ref{eq:cost_inequlality}) can be rewritten as
\begin{equation*}
	\EX[C^{\pi^*}(T)] - T\mc{P}({\lambda}) 	\geq  T \left(\mathcal{P}(\bar{\lambda})  - \mathcal{P}(\lambda) \right) + T C_B \big( \lambda-\bar{\lambda} \big)
\end{equation*}
Since $\mathcal{P}$ is a linear optimization problem with positive costs, $\mathcal{P}(\lambda)$ is convex in $\lambda$ (in fact, it is a convex piece-wise linear function \cite{bt_optimization_book}).
Since $\lambda \geq \bar{\lambda}$, we can use convexity to write $\mathcal{P}(\bar{\lambda})  \geq \mathcal{P}(\lambda)  - (\lambda - \bar{\lambda})  {d\mathcal{P}(\lambda)}/{d\lambda}$. Hence,
\begin{equation}\label{eq:cost_inequality_single}
	\EX[C^{\pi^*}(T)] - T\mc{P}(\lambda)  \geq T(\lambda-\bar{\lambda})\left[C_B - \frac{d\mathcal{P}(\lambda)}{d\lambda} \right].
\end{equation}

Note that ${d\mathcal{P}(\lambda)}/{d\lambda}$ is the minimum per-traffic marginal cost required to send an infinitesimally small additional traffic at $\lambda$. Recall that $\{ \mu^{stat}_{ij} : {(i,j)\in E} \}$ is the minimum cost flow to support the volume of $\lambda$. Hence, the quantity ${d\mathcal{P}(\lambda)}/{d\lambda}$ is the marginal cost required to send a small additional traffic on top of this flow. To send the additional traffic, we can increase the flow on edges with remaining capacities and/or reduce the flow on edges with non-zero flows, all while maintaining flow conservation. Accordingly, the resulting marginal cost can be calculated using a residual graph $G_R$ constructed as follows. The residual graph $G_R$ has the same set of vertices as $G$. Further, for each edge $(i,j)\in E$ in $G$, if $\mu_{ij}^{stat}<\mu_{ij}^{max}$, we add this edge to $G_R$ with the same original cost $c_{ij}$. Also, if $\mu_{ij}^{stat}>0$, we add the backward edge $(j,i)$ to $G_R$ with the negative cost $-c_{ij}$. The per-unit marginal cost to send an infinitesimally small additional traffic ${d\mathcal{P}(\lambda)}/{d\lambda}$ is equal to the cost of least expensive acyclic path on $G_R$. 

Pick the design parameter $C_L$ to be the most expensive acyclic path on the original network while incurring negative costs on backward edges. Notice that this choice of $C_L$ is finite and only depends on edge costs and the original network topology. By definition of $C_L$, it must be greater than the minimum per-traffic cost required to send an infinitesimal traffic on the residual graph. Therefore, we have ${d\mathcal{P}(\lambda)}/{d\lambda} \leq C_L.$
Plugging this in (\ref{eq:cost_inequality_single}), and using the fact that $\lambda\geq\bar{\lambda}$, we can conclude that, when $C_B \geq C_L$,
\begin{equation*}
	\EX[C^{\pi^*}(T)] - T\sum_{(i,j)\in \mathcal{E}} \mu^{stat}_{ij}c_{ij} = \EX[C^{\pi^*}(T)] - T\mc{P}(\lambda) \geq 0.
\end{equation*}
This is the desired result for the special case of networks with a single source-destination pair. A general proof for the multi-commodity case is given in Appendix \ref{proof:static_lower_bound_multi}.
\hfill \IEEEQEDhere

Note that the theorem and its proof suggest that a good choice of $C_B$ should be larger than $C_L$. Under this choice, minimizing $C^\pi(T)$ ensures network stability. Otherwise, a policy may be able to achieve lower costs by leaving more packets in the buffers at the end of time horizon, and fail to stabilize the network. Since our goal is to design stabilizing policies, we assume that $C_B \geq C_L$ for the rest of the paper.
Further, note that even though the solution to the optimization problem $\mc{P}$ lower bounds the cost, a policy that chooses transmissions equal to $\bs{\mu}^{stat}$ at all time $t$ is not optimal. This is because a portion of the arrived packets will remain in the queues due to the randomness in arrivals. Hence, including the queue backlog penalty, such a policy's total cost will be greater than $T \sum_{(i,j)\in \mc{E}} \sum_{k\in\mc{N}} \mu_{ijk}^{stat}c_{ij}$. Moreover, since the  arrival rate and costs are unknown, it is impossible to calculate the solution to $\mc{P}$. We only use the solution to  $\mc{P}$ as a bound to analyze our policy's regret and our policy will not need to solve this optimization problem. Finally, as a direct implication of Theorem \ref{thm:static_lower_bound}, we can express the regret of any policy as follows.
\begin{corollary}\label{corollary:static} For $C_B \geq C_L$, we have
	\begin{multline*}
		R^\pi(T) \leq \sum_{t=1}^T \sum_{(i,j)\in \mc{E}} \sum_{k\in\mc{N}} \Ex\left[\mu^\pi_{ijk}(t) - \mu^{stat}_{ijk}\right]c_{ij} \\
		+ C_B\sum_{i\in \mc{N}} \sum_{k\in\mc{N}} \Ex \left[Q_{ik}^\pi(T)\right].
	\end{multline*}
\end{corollary}
\begin{IEEEproof}[Proof of Corollary \ref{corollary:static}]
	The result directly follows from the definition of regret, the cost bound (\ref{eq:multi_planned_costs}), and Theorem \ref{thm:static_lower_bound}.
\end{IEEEproof}

It can be seen from Corollary \ref{corollary:static} that any policy with sub-linear regret will also be rate stable. Hence, our objective of finding a policy that has sub-linear regret inherently ensures network stability.
We discuss our proposed Drift Plus Optimistic Penalty policy in the next section.

\section{Drift Plus Optimistic Penalty Policy}\label{sec:algorithm}

We use the technique of drift-plus-penalty minimization from \cite{neely_known_costs} to derive our policy. We first define the Lyapunov function under a given policy $\pi$ at time $t$ as
\begin{equation*}
	\Phi^\pi(t) := \frac{1}{2} \sum_{i\in \mc{N}}\sum_{k\in \mc{N}} Q_{ik}^\pi(t)^2.
\end{equation*}
We denote the Lyapunov drift at time $t$ as $\Delta\Phi^\pi(t) := \Phi^\pi(t+1) - \Phi^\pi(t)$. Further, given queue backlogs, we can define the Lyapunov drift-plus-penalty at time $t$ as
$$
	L^\pi(t) := \EX \Bigg[ \Delta\Phi^\pi(t) + \nu \sum_{(i,j)\in \mc{E}} \sum_{k \in \mc{N}}  \mu_{ijk}^\pi(t)c_{ij} \;\bigg| \; \bs{Q}^\pi(t) \Bigg]
$$
where, $\nu$ is a tuning parameter that will be used to tune the cost-backlog trade-off, where the penalty term on the right represents the total transmission cost at time $t$.

The idea behind drift-plus-penalty minimization technique is to greedily minimize $L^\pi(t)$ at each time-slot $t$. However, we cannot do this directly in our setting  as the edge costs $c_{ij}$'s are unknown. Instead, we have to estimate the costs using past observations and make our decisions based on these estimates. Moreover, since we get observations for only the edges we select, we face an exploration-exploitation trade-off. We have to simultaneously exploit past observations and explore less observed edges to improve their cost estimates. This motivates us to use the idea of \emph{optimism in face of uncertainty} from the multi-arm bandits literature \cite{auer_bandits, bianchi_bandits}.

For edge $(i,j)\in\mc{E}$, recall that $e_{ij}(t)\in\{0,1\}$ is the indicator variable that tracks whether cost was observed at time $t$, and $\widetilde{c}_{ij}(t)$ is the observed cost. Let $N_{ij}(t) := \sum_{\tau<t}e_{ij}(\tau)$ be the number of observations received until time $t$ for edge $(i,j)\in\mc{E}$. Denote by $\bar{c}_{ij}(t)$ the average cost observation until time $t$,
$
	\bar{c}_{ij}(t) := \frac{1}{N_{ij}(t)}\sum_{\tau=0}^{t-1} \widetilde{c}_{ij}(\tau) e_{ij}(\tau).
$
We define the lower-confidence-bound estimate of the cost of edge $(i,j)\in\mc{E}$ at time $t$ as
\begin{equation}\label{eq:cost_estimate}
	 \hat{c}_{ij}(t) :=  \bar{c}_{ij}(t) - \sqrt{\frac{\beta \log (t/\delta)}{N_{ij}(t)}}
\end{equation}
where, $\delta \in (0,1)$ and $\beta \geq 0$ are tuning parameters. Note that this is an optimistic estimate of the costs, where the term $\sqrt{{\beta \log (t/\delta)}/{N_{ij}(t)}}$ adds a bias in favor of under-explored edges whose $N_{ij}(t)$ values are small. Now, as the true costs are unknown, we instead use these optimistic estimates in the drift-plus-penalty expression. Formally, we use the new drift-plus-optimistic-penalty expression $\hat{L}^\pi(t)$ defined as
\begin{equation*}
	\hat{L}^\pi(t):= \EX_{\:| Q^\pi, \hat{c}} \biggl[ \Delta\Phi^\pi(t) + \nu \sum_{(i,j)\in \mc{E}} \sum_{k\in\mc{N}} \mu_{ijk}^\pi(t) \hat{c}_{ij}(t) \biggr]
\end{equation*}
where, $\EX_{\:| Q^\pi, \hat{c}}[\cdot] := \EX[\cdot \;|\; \bs{Q}^{\pi}(t), \{ \hat{c}_{ij}(t) : {(i,j)\in\mc{E}}\} ]$ is the conditional expectation given queue backlogs and cost estimates.
From the queue evolution dynamics (\ref{eq:multi_queue_evol_i}), and from the fact that $([q-b]^+ +a)^2 \leq q^2 + b^2 + a^2 + 2q(a-b)$, we obtain the following bound on drift-plus-optimistic-penalty.
\begin{multline*}
	\hat{L}^\pi(t) \leq B + \sum_{i\in\mc{N}} \sum_{k\in\mc{N}} \lambda_{ik} Q_{ik}^\pi(t) \\
	+  \EX_{\:| Q^\pi, \hat{c}}\biggl[  \sum_{(i,j)\in \mc{E}} \sum_{k\in\mc{N}} \mu_{ijk}^\pi(t)\left(Q^\pi_{jk}(t) - Q_{ik}^\pi(t) + \nu \hat{c}_{ij}(t)  \right) \biggr]
\end{multline*}
where, $B := \frac{1}{2} \sum_{i\in \mc{N}} \sum_{k\in \mc{N}} [(\sum_{j\in \mc{N}_i}\mu_{ij}^{max})^2 + \EX[a^2_{ik}(t)] + (\sum_{j:i\in \mc{N}_j}\mu_{ji}^{max})^2 + 2\lambda_{ik} \sum_{j:i\in\mc{N}_j} \mu_{ji}^{max}]$ is a constant that depends only on edge capacities, arrival rates, and the second moment of arrivals.

Now, we derive the Drift Plus Optimistic Penalty (DPOP) policy $\pi_{D}$ by minimizing this bound as shown below. Define the set of all feasible transmissions as
$\mc{M} := \{\bs\mu \geq 0: \forall(i,j)\in \mc{E},\; \sum_{k\in\mc{N}} \mu_{ijk} \leq \mu^{max}_{ij}\}.$
Given the backlogs $\bs{Q}^{\pi_D}(t)$ and the lower-confidence-bound cost estimates $\{\hat{c}_{ij}(t) : (i,j)\in\mc{E}\}$ at time $t$, the policy $\pi_D$ picks edge transmission values
$\bs{\mu}^{\pi_{D}}(t)$ 
by greedily minimizing the bound on $	\hat{L}^\pi(t)$. Ignoring the uncontrollable terms, this simplifies to
\begin{multline}\label{eq:transmission_decision}
		\bs{\mu}^{\pi_{D}}(t) = \arg \max_{\bs{\mu} \in \mc{M}} \sum_{(i,j)\in \mc{E}}  \sum_{k\in\mc{N}}  \mu_{ijk} \Big( Q_{ik}^{\pi_D}(t) \\
		- Q_{jk}^{\pi_D}(t) - \nu \hat{c}_{ij}(t) \Big).
\end{multline}

We present the DPOP policy in Algorithm \ref{alg:dpop}.
We start with an initial exploration phase at $t=0$ before any packets arrive. In this phase, we send a dummy packet on each edge (line 1) to receive an initial cost observation (line 2). Using this observation, we initialize the average cost estimates $\bar{c}_{ij}(t)$ (line 3) and $N_{ij}(t)$ (line 4) for $t=1$. New packets arrive starting from $t=1$. At each time $t\in \{1,...,T\}$, we calculate the optimistic estimates $\hat{c}_{ij}(t)$ (line 5) and make transmission decisions  $\mu^{\pi_D}_{ijk}(t)$ by greedily minimizing the bound on the Lyapunov drift-plus-optimistic-penalty (line 6). It is possible that for some edges, their neighboring nodes do not have sufficient packets to support the planned transmissions. On those edges, we send dummy packets such that the total transmission is no more than the planned transmission (line 7). This allows us to get observations for all the chosen edges. Note that the dummy packets do not contribute to the queue evolution but they do incur transmission costs just like normal packets. We then perform the planned transmissions, which updates the queue backlogs (line 8). We receive noisy cost observations for each edge on which we transmitted packets (line 9). Finally, we update the cost averages $\bar{c}_{ij}(t)$ (line 10) and the number of observations $N_{ij}(t)$ (line 11).

\begin{algorithm}[htbp]
	\caption{Drift Plus Optimistic Penalty (DPOP) Algorithm.}\label{alg:dpop}
	\begin{algorithmic}
		\STATE \textbf{Input:} Network $G$, Parameters $\beta$, $\delta$, and $\nu$.
		\STATE \textbf{at $t=0$ do}
		\algoline{1}{Send a dummy packet on each edge, $\mu^{\pi_{D}}_{ijk}(0) = 1, \;\forall ij$.}
		\algoline{2}{Observe noisy costs $\widetilde{c}_{ij}(0), \; \forall(i,j)\in \mc{E}$.}
		\algoline{3}{Set $ \bar{c}_{ij}(1) =  \widetilde{c}_{ij}(0), \; \forall(i,j)\in \mc{E}$.}
		\algoline{4}{Set $N_{ij}(1) = 1, \; \forall(i,j)\in \mc{E}$.}
		\FOR{$t=1,2,...,T$}
		\algoline{5}{Calculate cost estimates $ \{\hat{c}_{ij}(t)\}_{(i,j)\in \mc{E}}$ using (\ref{eq:cost_estimate}).}
		\algoline{6}{Pick transmissions $\bs{\mu}^{\pi_{D}}(t)$ according to (\ref{eq:transmission_decision}).}
		\algoline{7}{Send dummy packets on each edge $(i,j)\in \mc{E}$ where $\mu^{\pi_{D}}_{ijk}(t) > 0$ but queue $Q^{\pi_D}_{ik}(t)$ does not have sufficient packets to support the planned transmission $\mu^{\pi_{D}}_{ijk}(t) $.}
		\algoline{8}{Update queue lengths according to (\ref{eq:multi_queue_evol_i}).}
		\algoline{9}{Observe noisy costs $\widetilde{c}_{ij}(t)$ for all $(i,j): e_{ij}(t) = 1$.}
		\algoline{10}{$\forall(i,j) :e_{ij}(t) = 1$, update average costs $\bar{c}_{ij}(t+1)$
			\begin{equation*}
				\bar{c}_{ij}(t+1) =  \bar{c}_{ij}(t)+\frac{\widetilde{c}_{ij}(t) -   \bar{c}_{ij}(t)}{N_{ij}(t)+1}.
			\end{equation*}\vspace{-6pt}}
		\algoline{11}{$\forall(i,j): e_{ij}(t) = 1,\; N_{ij}(t+1) = N_{ij}(t)+1$.}
		\ENDFOR
	\end{algorithmic}
\end{algorithm}


\section{Regret Analysis}\label{sec:regret_analysis}
In this section, we analyze the performance of the DPOP policy $\pi_D$ and derive an upper-bound on its regret. We formally present the regret performance in the following theorem.

\begin{theorem}\label{thm:regret_theorem} \emph{(DPOP Regret)} For $C_B \geq C_L$, the DPOP policy $\pi_D$ achieves a sub-linear regret of
	\begin{equation*}
		R^{\pi_D}(T) = O\left(\sqrt{T}\log T\right)
	\end{equation*}
	with parameters $\beta > 4\sigma^2$, $\delta = T^{-{2\sigma^2}/{\beta}}$, and $\nu = \sqrt{T}$.
\end{theorem}
\emph{Proof of Theorem \ref{thm:regret_theorem}:} The proof outline is as follows. We first decompose $\pi_D$'s regret into four components in Lemma \ref{lemma:regret_decomp}. We then bound each of these components individually in Propositions \ref{prop:exploration_regret} to \ref{prop:queueing_regret}.
We first define event $A$ as the event that the true edge costs $c_{ij}$'s are within the confidence interval of our estimates $ \bar{c}_{ij}(t)$'s at all slots $t$ and at all edges.
\begin{equation*}
	A := \left\{\forall t, \forall (i,j), \; |c_{ij} -  \bar{c}_{ij}(t)| \leq \sqrt{\frac{\beta\log (t/\delta)}{N_{ij}(t)}} \right\}.
\end{equation*}

Denote the compliment of event $A$ by $\bar{A}$. Now, in Lemma \ref{lemma:regret_decomp}, we decompose the regret into four components $R_1^{\pi_D}(T), ..., R_4^{\pi_D}(T)$ defined as follows.
	\begin{align*}
		R^{\pi_D}_1(T) &:= \sum_{t=1}^T\sum_{(i,j)\in \mathcal{E}} \sum_{k \in \mc{N}} \Ex \left[ \mu^{\pi_D}_{ijk}(t)(c_{ij} - \hat{c}_{ij}(t)) \;|\; A \right], \\
		R^{\pi_D}_2(T) &:= T|\mc{N}| \sum_{(i,j)\in \mathcal{E}}\mu_{ij}^{max}C_{max} \PR[\bar{A}], \\
		R^{\pi_D}_3(T) &:= \sum_{t=1}^T\sum_{(i,j)\in \mathcal{E}} \sum_{k \in \mc{N}} \EX [(\mu^{\pi_D}_{ijk}(t) - \mu_{ijk}^{stat}) \hat{c}_{ij}(t) \;|\; A] \Pb[A], \\
		R^{\pi_D}_4(T) &:= C_B \sum_{i\in \mathcal{N}} \sum_{k \in \mc{N}} \EX[Q^{\pi_D}_{ik}(T)].
	\end{align*}

The first component $R_1^{\pi_D}(T)$ corresponds to the gap between the true and estimated costs. The second component $R_2^{\pi_D}(T)$ captures the probability that the true cost is outside the confidence interval of our estimate. The third component $R_3^{\pi_D}(T)$ corresponds to the gap between policy's transmission cost and the static cost. Finally, the component $R_4^{\pi_D}(T)$ corresponds to the backlog cost.

\begin{lemma}\label{lemma:regret_decomp} \emph{(Regret Decomposition)} The regret $R^{\pi_D}(T)$ of the DPOP policy $\pi_D$  can be decomposed as
	\begin{equation*}
		R^{\pi_D}(T) \leq R^{\pi_D}_1(T) + R^{\pi_D}_2(T) + R^{\pi_D}_3(T)  + R^{\pi_D}_4(T).
	\end{equation*}
\end{lemma}

We prove Lemma \ref{lemma:regret_decomp} in Appendix \ref{proof:regret_decomp}. The proof involves using the regret upper bound from Corollary \ref{corollary:static} and conditioning the regret on events $A$ and $\bar{A}$. We ignore the cost of exploration phase at $t=0$ as this only adds a constant term to the regret. We bound the components in the following propositions.
\begin{itemize}
	\itemindent=-8pt
	\item \begin{proposition}\label{prop:exploration_regret}
		$R^{\pi_D}_1(T)= O(\sqrt{T}\log T).$
	\end{proposition}

	\item \begin{proposition}\label{prop:regret_3_bound}
		$R^{\pi_D}_2(T) = O(1)$ with $\delta = T^{\frac{-2\sigma^2}{\beta}}$\hspace{-1.25mm}, $\beta > 4\sigma^2$\hspace{-1.25mm}.
	\end{proposition}

	\item \begin{proposition}\label{prop:cost_gap_to_stat}
		$R^{\pi_D}_3(T)= O(\sqrt{T})$  with $\nu=\sqrt{T}$.
	\end{proposition}

	\item \begin{proposition}\label{prop:queueing_regret}
		$R^{\pi_D}_4(T)= O(\sqrt{T \log T})$ with $\nu=\sqrt{T}$.
	\end{proposition}
\end{itemize}

We prove Propositions \ref{prop:exploration_regret} to \ref{prop:queueing_regret} in Appendices \ref{proof:exploration_regret} to \ref{proof:queueing_regret} respectively. Combining Lemma \ref{lemma:regret_decomp} and Propositions \ref{prop:exploration_regret} to \ref{prop:queueing_regret}, we can see that the overall regret of the policy is $O(\sqrt{T}\log T)$. This proves Theorem \ref{thm:regret_theorem} and concludes the regret analysis. \hfill\IEEEQEDhere

Theorem \ref{thm:regret_theorem} shows that the DPOP policy $\pi_D$ achieves sub-linear regret. Moreover, we can see from Proposition \ref{prop:queueing_regret} that
$$\lim_{T\rightarrow\infty} \frac{1}{T}\sum_{i\in \mathcal{N}} \sum_{k \in \mc{N}} \EX[Q^{\pi_D}_{ik}(T)] = \lim_{T\rightarrow\infty} \frac{R^{\pi_D}_4(T)}{T C_B} = 0.$$
This shows that the proposed DPOP policy is also rate stable. 

In Theorem \ref{thm:regret_theorem}, we expressed the regret as a function of $T$ up to constant factors. We now take a closer look at the multiplicative constants to understand the policy's dependency on the terminal backlog cost $C_B$. From (\ref{eq:R_3_dep_on_CB}) and (\ref{eq:R_4_dep_on_CB}) in proofs of Propositions \ref{prop:cost_gap_to_stat} and \ref{prop:queueing_regret}, we can see that the optimal choice of $\nu$ is $\nu = O (\sqrt{T/\max (C_B, 1)} )$, which achieves a regret of $ O (\min(\sqrt{C_B}, C_B) \cdot \sqrt{T}\log T )$ when $C_B \geq C_L$.
The inverse dependency of $\nu$ on $C_B$ highlights the DPOP policy's behavior as the terminal backlog cost $C_B$ varies. When  the terminal backlog cost $C_B$ is high, the value of $\nu$ is small, which causes the decision rule (\ref{eq:transmission_decision}) to prioritize balancing the queue differentials, and in turn reducing the terminal queue backlog. On the other hand, when the terminal backlog cost $C_B$ is low, the value of $\nu$ is large, which causes the policy to prioritize lowering the transmission costs.


Further, Theorem \ref{thm:regret_theorem} and its choice of parameters is valid only under the assumption that $C_B \geq C_L$. This assumption allowed us to lower bound the best policy's cost in terms of the static optimal cost in Theorem \ref{thm:static_lower_bound}. Using this, we could upper bound the regret  as
$R^\pi(T) \leq \sum_t \sum_{(i,j)} \sum_{k} \Ex[\mu^\pi_{ijk}(t) - \mu^{stat}_{ijk} ]c_{ij}
+ C_B\sum_{i} \sum_{k} \Ex \left[Q_{ik}^\pi(T)\right].$
We then chose the parameters  $\nu$, $\delta$, and $\beta$  in Theorem \ref{thm:regret_theorem}  specifically to optimize this upper bound.
However, when $C_B < C_L$, this upper bound is no longer valid and hence, Theorem \ref{thm:regret_theorem}'s parameter choice does not result in a sub-linear regret.
As mentioned before, when $C_B < C_L$, the best policy may be able to achieve a lower cost by leaving packets behind in the network.
Hence, in order to achieve sub-linear regret, the DPOP policy may also need to ignore queue stability.
This requires a separate analysis to check if such a choice of parameters even exists. Further, this analysis must also find a new lower bound on the best policy's cost since Theorem \ref{thm:static_lower_bound} may not be true anymore. As our objective is to find stabilizing policies, such an analysis is beyond the scope of this paper. Instead, we conduct simulations in Section \ref{sec:small_CB_experiment} with $C_B < C_L$  to study the behavior of the best policy and the DPOP policy with Theorem \ref{thm:regret_theorem}'s choice of parameters.




\section{Simulation Results}\label{sec:results}
\subsection{Single-Commodity Network Simulation}\label{sec:single-commodity-sims}
\begin{figure}[htbp]
	\centering
	\includegraphics[width=0.3\textwidth]{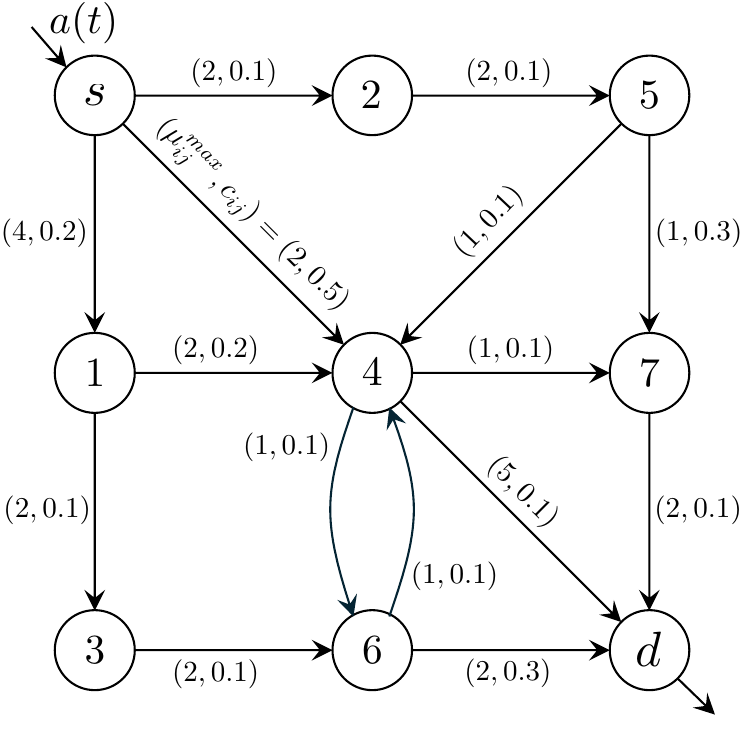}
	\caption{Single-commodity network showing $(\mu_{ij}^{max}, c_{ij})$.}\label{fig:topology}
\end{figure}

We evaluate\footnote{\url{https://github.com/SathwikChadaga/Optimistic-DPP-Improved-Regret}} our policy on a single-commodity queueing network with 9 nodes and 15 edges shown in Fig. \ref{fig:topology}. The edge capacities and transmission costs are shown as tuples $(\mu_{ij}^{max}, c_{ij})$ marked on their respective edges in the figure. The network has a single source-destination pair, marked as $s$ and $d$ respectively in the figure. For this network, we can calculate the stability region using (\ref{eq:capacity_region_multi}) to be $\lambda \in [0,8]$. New packets arrive at $s$ according to a Poisson process with rate within this interval. All cost observations are corrupted by independent random variables uniformly distributed in $[-\sigma, \sigma]$.

\subsubsection{Queue Backlog and Transmission Cost Performance} We simulate the DPOP policy for $T=10000$, $\lambda = 4$, and $\sigma^2 = 0.05$. We simulate the policy 10000 times and take average of these runs to estimate expected values. We use Theorem \ref{thm:regret_theorem} to choose tuning parameters $\beta = 4.5\sigma^2$, $\delta = T^{-{2\sigma^2}/{\beta}}$, and $\nu = \sqrt{T}$.
This choice of tuning parameters requires the knowledge of $T$, which may not be available in some practical scenarios. Hence, we also simulate the policy for the case of unknown $T$ using the doubling trick \cite{doubling}. Here, we pick the tuning parameters according to an estimated time horizon $\hat{T}$ (initialized as $\hat{T}=2$), which we double whenever the current time-slot $t$ crosses the current estimate, i.e. $\hat{T}\leftarrow 2\hat{T}$ if $t>\hat{T}$. To benchmark our policy, we also simulate an oracle policy that has access to the true transmission costs. The oracle policy is similar to the DPOP policy described in Algorithm \ref{alg:dpop} except that in line 6, it uses the true costs $c_{ij}$'s instead of the optimistic estimates $\hat{c}_{ij}(t)$'s, and it skips lines 9, 10, and 11 as it already knows the true edge costs. Finally, we also simulate a bandit shortest path (Bandit SP) policy that naively sends each packet through the least expensive path in terms of the UCB cost estimates. This policy completely disregards the queueing backlogs, and greedily learns and minimizes the transmission costs.

\begin{figure}[htbp]
	\centering
	\includegraphics[width=0.95\linewidth]{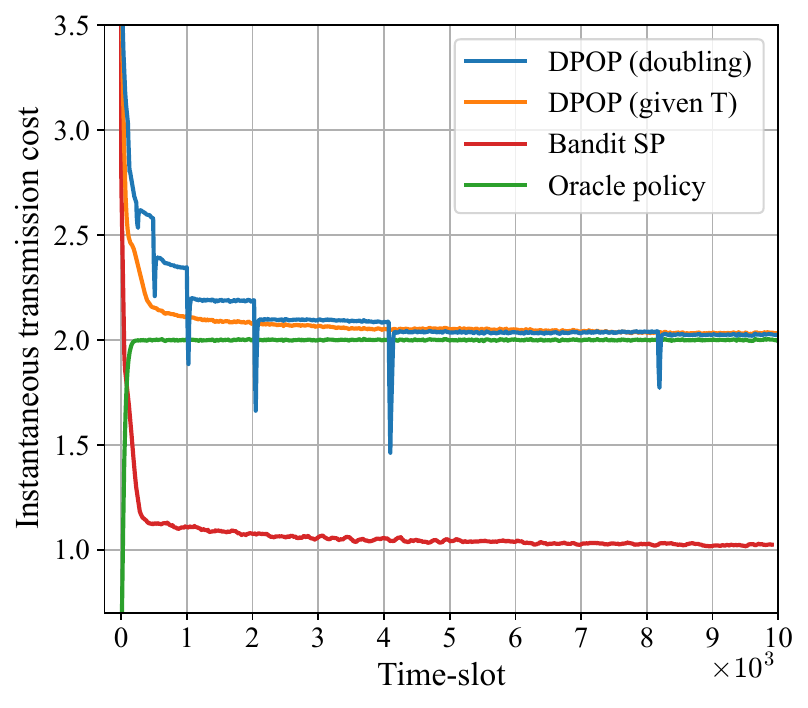}
	\caption{Transmission cost $\sum_{(i,j)\in \mathcal{E}}\Ex[\mu_{ij}^{\pi}(t)]c_{ij}$.}\label{fig:costs}
\end{figure}

Fig. \ref{fig:costs} shows the plot of resulting total transmission costs $\sum_{(i,j)\in \mathcal{E}}\Ex[\mu_{ij}^{\pi}(t)]c_{ij}$ as a function of time $t$.
The figure has four plots corresponding to the DPOP policy with doubling trick (unknown $T$), DPOP policy with known $T$, bandit shortest path policy, and the oracle policy.
From this figure, we can see that the DPOP policy's cost is initially large, since at this time, it is still exploring and learning the edge costs. As $t$ increases, the policy's estimates improve, and hence its cost reduces and eventually approaches the oracle policy.
This shows the policy's learning process and its ability to explore and learn the low cost routes. Further, the bandit shortest path policy has the lowest transmission cost since it greedily chooses the shortest path even if that path is backlogged. This will impact its backlog performance as discussed next.

\begin{figure}[htbp]
	\centering
	\includegraphics[width=\linewidth]{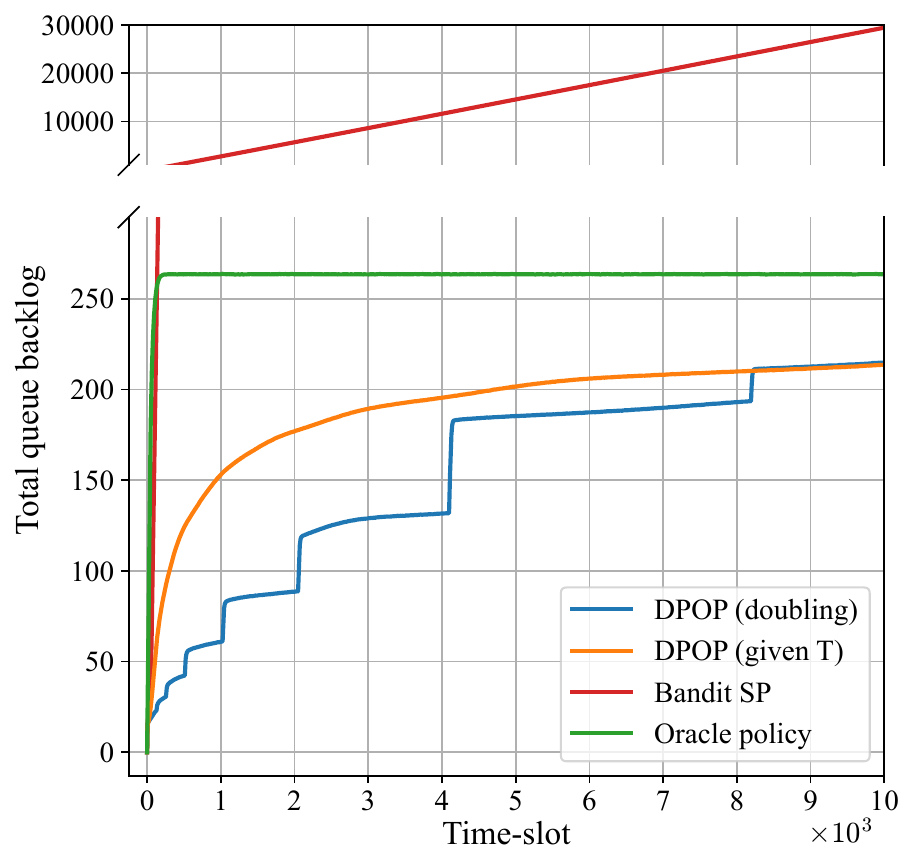}
	\caption{Total queue backlog $\sum_{i\in \mathcal{N}} \Ex[Q_i^{\pi}(t)]$.}\label{fig:backlogs}
\end{figure}

Fig. \ref{fig:backlogs} shows the plot of total queue backlogs $\sum_{i\in \mathcal{N}} \Ex[Q_i^{\pi}(t)]$ as a function of time $t$.
From this figure, we can see that DPOP policy's queue backlog does not grow indefinitely, which demonstrates its ability to stabilize the network. Further, compared to the oracle policy, we can see that the DPOP policy has a lower queue backlog but starts with higher transmission costs. This is due to the exploratory bias in DPOP's decision making. Indeed, as part of the exploration, it initially uses higher cost paths and hence reduces some queue backlogs. Additionally, we can see that the performance of DPOP policy with doubling (for unknown $T$) is close to the performance of DPOP policy with known $T$. This shows that the policy can also be used when the value of $T$ is not known in advance. Finally, we can see that the backlog of bandit shortest path policy grows indefinitely. This policy learns the least expensive path, which is $s, 2, 5, 4, d$, and greedily sends most packets through this route, apart from the occasional exploratory packets sent through other routes. However, the capacity of this path is 1. And since the arrival rate $\lambda = 4 > 1$, the queue backlogs in this path build up, making the network unstable. This highlights the importance of considering the queue backlogs when making transmission decisions.

\begin{figure}
	\centering
	\includegraphics[width=0.9\linewidth]{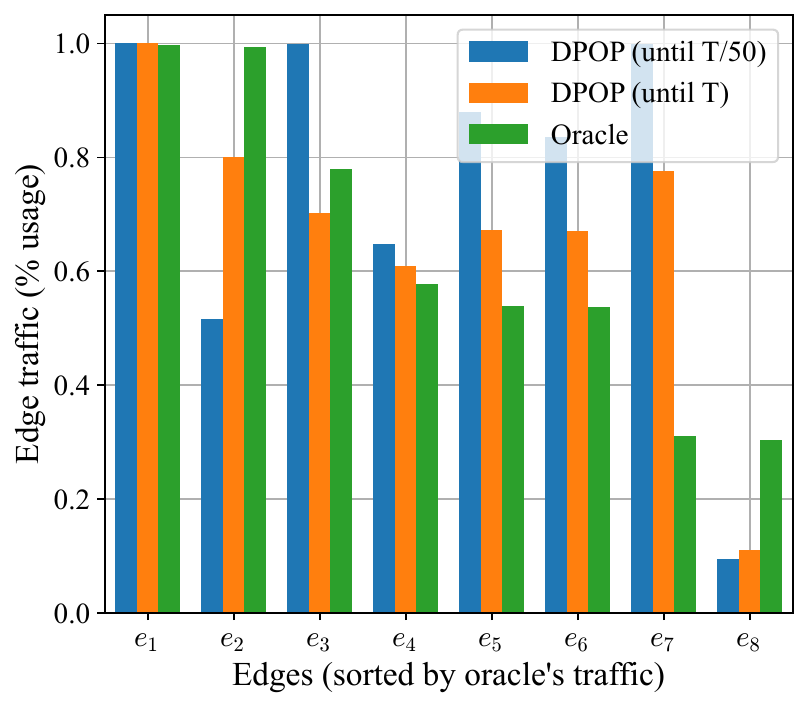}
	\caption{Edge utilization\hspace{1pt}$\frac{1}{t}\sum_{\tau=1}^t\hspace{-2pt} \Ex[\mu_{ij}^{\pi}(\tau)]/\mu_{ij}^{max}$.}\label{fig:utilization}
\end{figure}

Fig. \ref{fig:utilization} shows the plot of time-average edge utilization at various points of time during the experiment.
We calculate the edge utilization as $\frac{1}{t}\sum_{\tau=1}^t\Ex[\mu_{ij}^{\pi}(\tau)]/\mu_{ij}^{max}$, and plot these values for the top eight utilized edges. For each of the top eight utilized edges, we plot DPOP's average utilization up to $t=T/50$ and $t=T$. We also plot the average edge utilization of the oracle policy. The edges on x-axis are sorted according to oracle policy's utilization, showing good edges towards the left and bad edges towards the right. We can see that the DPOP policy initially allocates less traffic to good edges and more traffic to bad edges as a result of exploration. But eventually at $t=T$, the average traffic utilization gets closer to the oracle policy. This shows the DPOP policy's ability to learn good resource allocation strategies from noisy feedback.

\subsubsection{Regret Comparison for Varying Noise and Arrival Rates}
We evaluate the DPOP policy's regret for different values of $\lambda$ and $\sigma$. To calculate the regret, we use the upper bound from Corollary \ref{corollary:static}. We obtain policy costs $C^{\pi}(T)$ from simulations and calculate the solution to the static optimization problem $\mathcal{P}$ using SciPy's linear programming tool. Employing the upper bound, we calculate the regret $R^{\pi_D}(T)$ as the gap between policy cost $C^\pi(T)$ and $T$ times the static optimal cost.

\begin{figure}[htbp]
	\centering
	\includegraphics[width=0.9\linewidth]{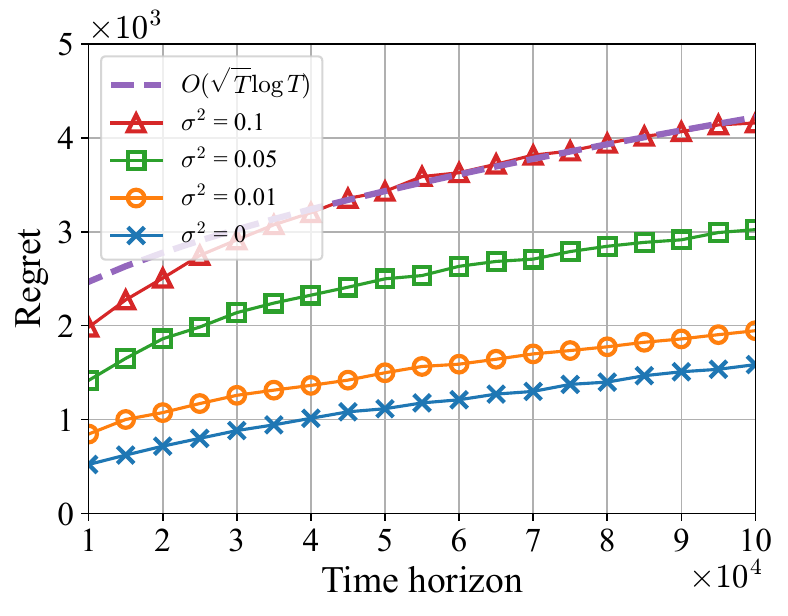}
	\caption{Regret for $\lambda = 2$.}\label{fig:regret_lambda_light}
\end{figure}

\begin{figure}[htbp]
	\centering
	\includegraphics[width=0.9\linewidth]{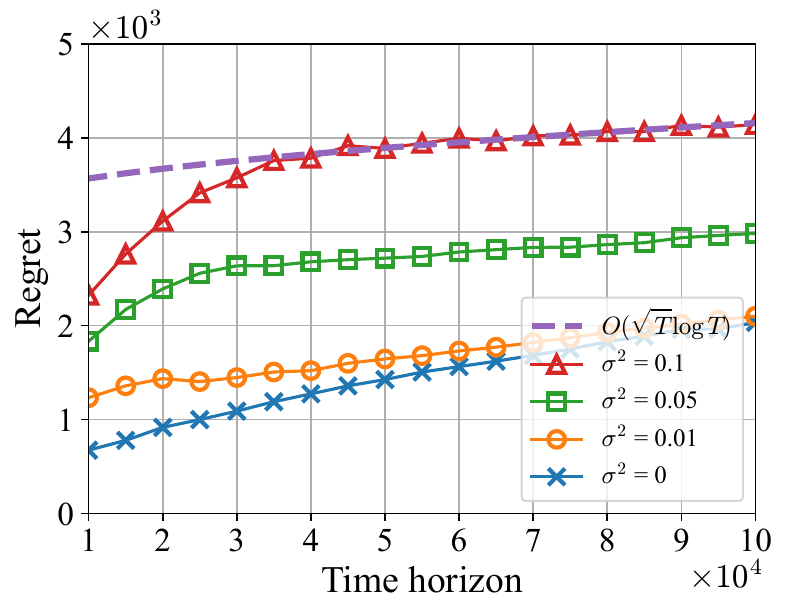}
	\caption{Regret for $\lambda = 3$.}\label{fig:regret_lambda_medium}
\end{figure}

\begin{figure}[htbp]
	\centering
	\includegraphics[width=0.9\linewidth]{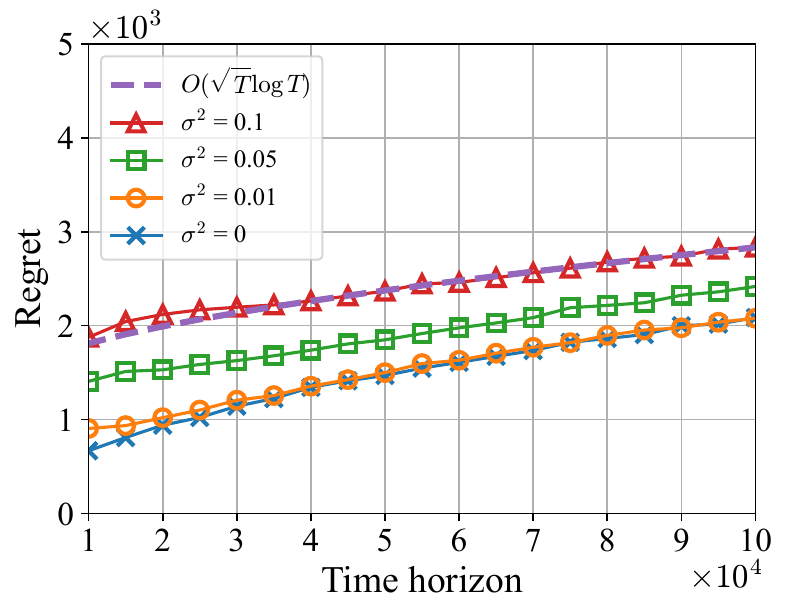}
	\caption{Regret for $\lambda = 4$.}\label{fig:regret_lambda_heavy}
\end{figure}

Figures \ref{fig:regret_lambda_light}-\ref{fig:regret_lambda_heavy} show the resulting regret plotted as a function of time horizon $T=\{10000,20000,...,100000\}$.
Specifically, figures \ref{fig:regret_lambda_light}, \ref{fig:regret_lambda_medium}, and \ref{fig:regret_lambda_heavy} show the regrets for $\lambda =2,3,$ and $4$ respectively. In each of these figures, we show the DPOP policy's regrets for different values of noise parameter $\sigma^2 = \{0,0.01, 0.05, 0.1\}$. We also plot a $O(\sqrt{T}\log T)$ curve fitted to the DPOP regret of $\sigma^2=0.1$ (we omit other values to avoid crowding). This demonstrates that the DPOP policy indeed has a sub-linear regret.

From the plots, we can see that the regret increases as the noise level increases, as expected.
Further, we can see that the regret for $\lambda=4$ is closer to the oracle policy (the case of $\sigma=0$), whereas regrets for $\lambda=2,3$ are higher than the oracle policy. This is because the rates $\lambda=2,3$ are farther from the network's max-flow of 8. Due to this light loading, the policies have more path choices requiring more exploration. Hence, the regret compared to the static optimal policy is large.
Finally, we clarify that the oracle policy (the case of $\sigma=0$) has a non-zero regret even though it knows the true costs because of the stochastic nature of arrivals. The random packet arrivals result in a non-zero gap between oracle policy's cost and the static lower bound.


\subsection{Multi-Commodity Network Simulation}\label{sec:multi-commodity-sims}
\begin{figure}[htbp]
	\centering\vspace*{-12pt}
	\includegraphics[width=0.45\textwidth]{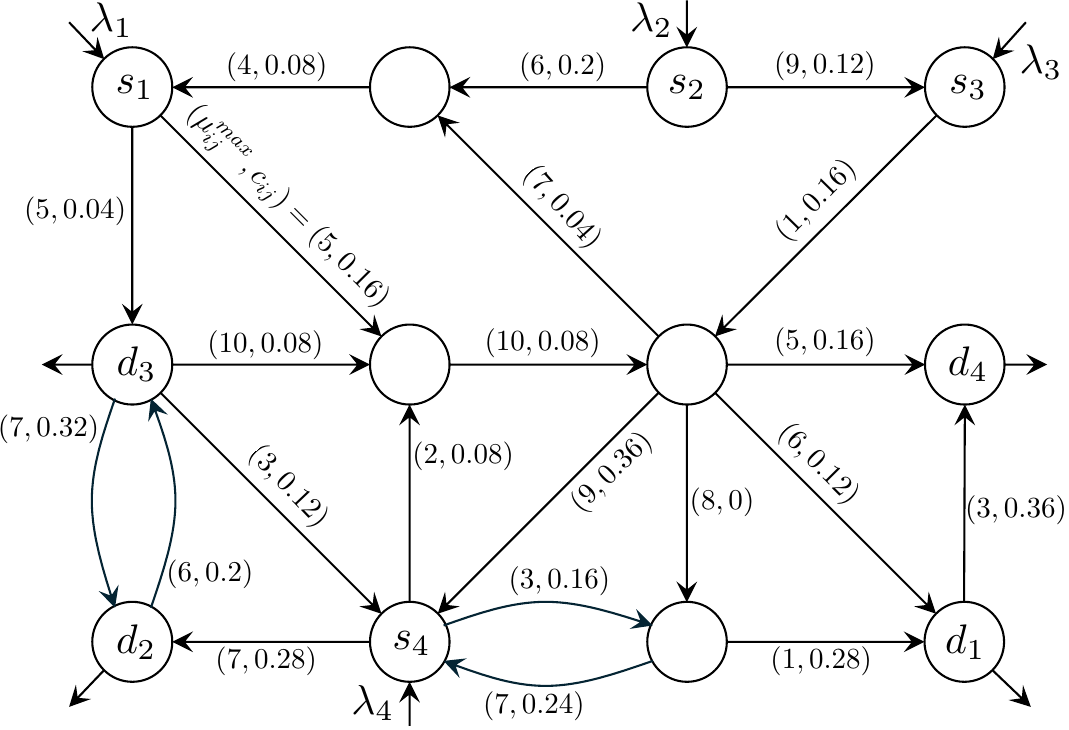}
	\caption{Multi-commodity network showing $(\mu_{ij}^{max}, c_{ij})$.}\label{fig:topology-multi}
\end{figure}

We evaluate our policy on a multi-commodity queueing network with 12 nodes and 22 edges shown in Fig. \ref{fig:topology-multi}. The edge capacities and transmission costs are shown as tuples $(\mu_{ij}^{max}, c_{ij})$ marked on their respective edges in the figure. This network serves 4 commodities and has 4 source-destination pairs corresponding to each commodity. The source nodes are marked as $\{s_1, s_2, s_3, s_4\}$, and the corresponding destination nodes are marked as $\{d_1, d_2, d_3, d_4\}$.
New packets arrive at source nodes according to a Poisson process with arrival rate vector $\bs{\lambda} := [\lambda_1, \lambda_2, \lambda_3, \lambda_4]$, where $\lambda_k$ is the arrival rate of commodity-$k$. Finally, all cost observations are corrupted by independent random variables uniformly distributed in $[-\sigma, \sigma]$.

\subsubsection{Queue Backlog and Transmission Cost Performance} We simulate the DPOP policy for time horizon $T=100000$, arrival rates $\bs{\lambda} = [2.5, 2.0, 0.5, 2.5]$, and noise variance $\sigma^2 = 0.1$. This arrival rate vector $\bs{\lambda}$ is within the capacity region for the chosen edge capacities. We use Theorem \ref{thm:regret_theorem} to choose tuning parameters $\beta = 4.5\sigma^2$, $\delta = T^{-{2\sigma^2}/{\beta}}$, and $\nu = \sqrt{T}$.
Similar to the single-commodity simulations, we benchmark our policy against an oracle policy, which is similar to the DPOP policy described in Algorithm \ref{alg:dpop} except that in line 6, it uses the true costs $c_{ij}$'s instead of the observed costs.

\begin{figure}[htbp]
	\centering
	\includegraphics[width=0.9\linewidth]{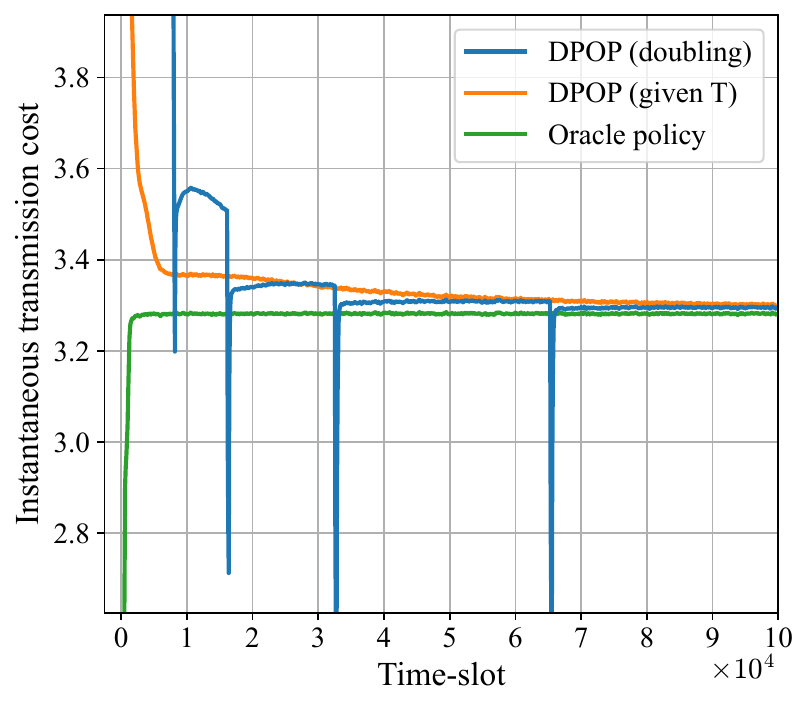}
	\caption{Transmit cost $\sum_{ij\in \mathcal{E}, k \in \mc{N}} \Ex[\mu_{ijk}^{\pi}(t)] c_{ij}$.}\label{fig:costs-multi}
\end{figure}

Fig. \ref{fig:costs-multi} shows the plot of resulting total transmission costs $\sum_{(i,j)\in \mathcal{E}} \sum_{k\in\mc{N}} \Ex[\mu_{ijk}^{\pi}(t)] c_{ij}$ as a function of time-slots $t$
for the DPOP policy with doubling trick (unknown $T$), DPOP policy with known $T$, and the oracle policy.
From this figure, we can see that the DPOP policy's cost is large initially and reduces later to eventually approach the oracle policy, which verifies the policy's ability to explore and learn low cost routes.

\begin{figure}[htbp]
	\centering
	\includegraphics[width=0.9\linewidth]{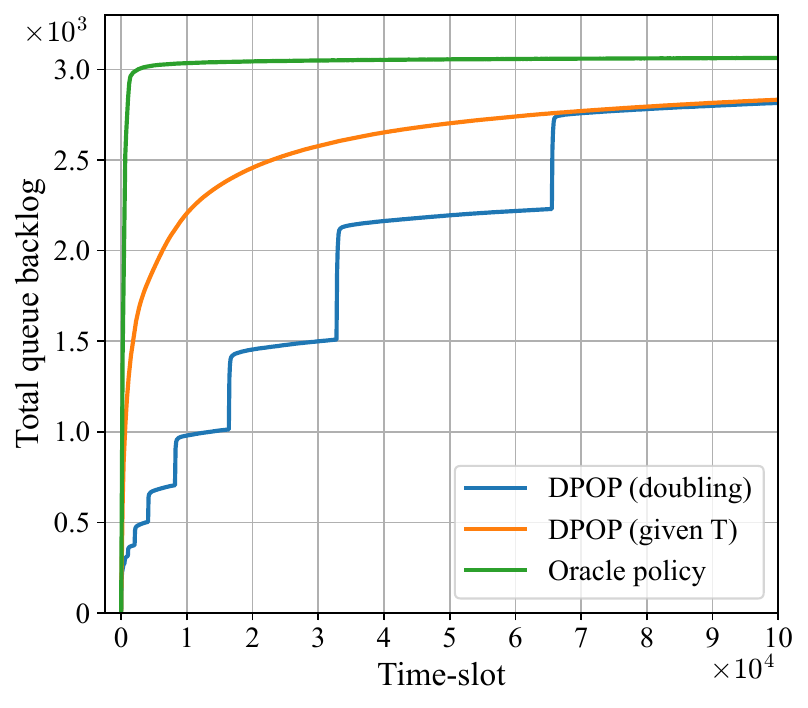}
	\caption{Total queue backlog $\sum_{i,k\in \mathcal{N}} \Ex[Q_{ik}^{\pi}(t)]$.}\label{fig:backlogs-multi}
\end{figure}

Fig. \ref{fig:backlogs-multi} shows the plot of total queue backlogs $\sum_{i\in \mathcal{N}} \sum_{k\in\mc{N}} \Ex[Q_{ik}^{\pi}(t)]$ as a function of time-slots $t$.
From this figure, we can see that the queue backlogs do not grow indefinitely, which demonstrates the DPOP policy's ability to stabilize the network.

\begin{figure}[htbp]
	\centering
	\includegraphics[width=0.9\linewidth]{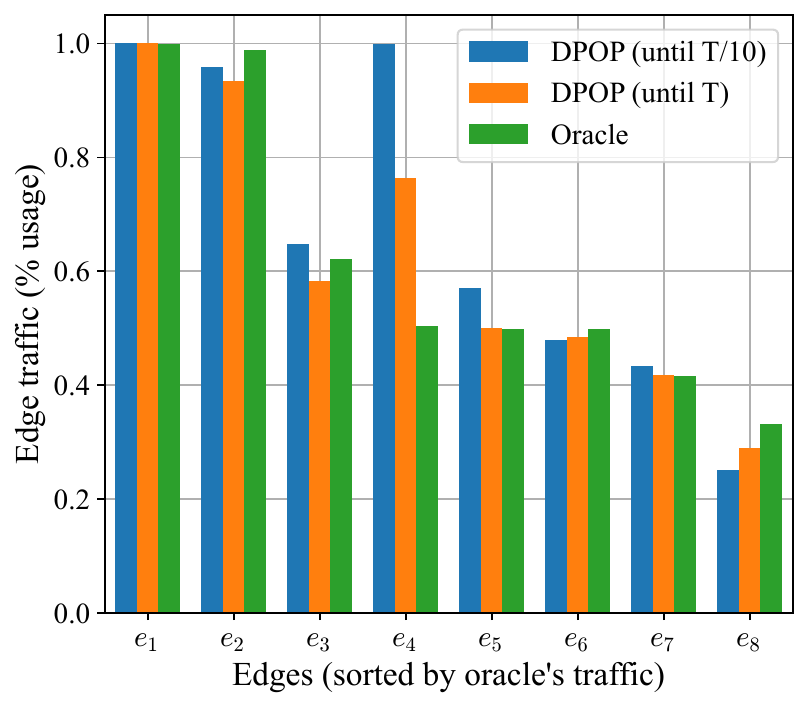}
	\caption{Edge usage $\frac{1}{t}\sum_{\tau=1,k}^t \Ex[\mu_{ijk}^{\pi}(\tau)] / \mu_{ij}^{max}$.}\label{fig:usage-multi}
\end{figure}

Finally, Fig. \ref{fig:usage-multi} shows the plot of time-average edge traffic for the top eight utilized edges.
We calculate the edge usage percentage as $\frac{1}{t}\sum_{\tau=1}^t \sum_{k\in\mc{N}} \Ex[\mu_{ijk}^{\pi}(\tau)] / \mu_{ij}^{max}$, and plot this value for $t=T/10$ and $t=T$. The edges on the x-axis are sorted according to oracle policy's traffic. Similar to single-commodity simulation, we can see that the DPOP policy is aggressive initially allocating large traffic rates to bad edges, eventually learning to allocate lesser traffic to bad edges.

\subsubsection{Regret Comparison for Varying Noise and Arrival Rates}
We evaluate the DPOP policy's regret for different values of $\bs{\lambda}$ and $\sigma$. Similar to the single-commodity simulations, we calculate the regret using the upper bound from Corollary \ref{corollary:static}. We obtain the policy costs $C^{\pi}(T)$ from simulations, calculate the solution to the static optimization problem $\mathcal{P}(\lambda)$ using SciPy's linear programming library, and calculate the regret $R^{\pi_D}(T)$ as the gap between $C^\pi(T)$ and $T$ times the static optimal cost.

\begin{figure}[htbp]
	\centering
	\includegraphics[width=0.9\linewidth]{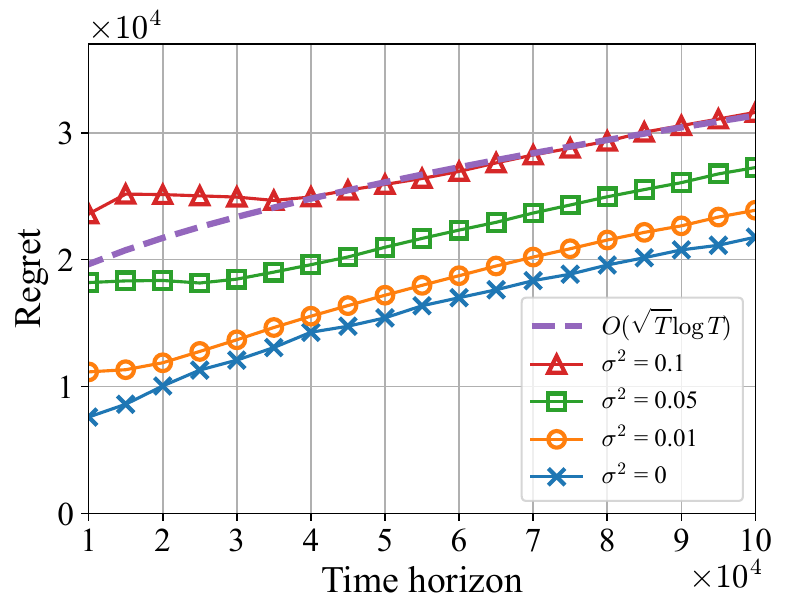}
	\caption{Regret for $\lambda = [0.83, 0.67, 0.17, 0.83]$.}\label{fig:regret-multi-light}
\end{figure}

\begin{figure}[htbp]
	\centering
	\includegraphics[width=0.9\linewidth]{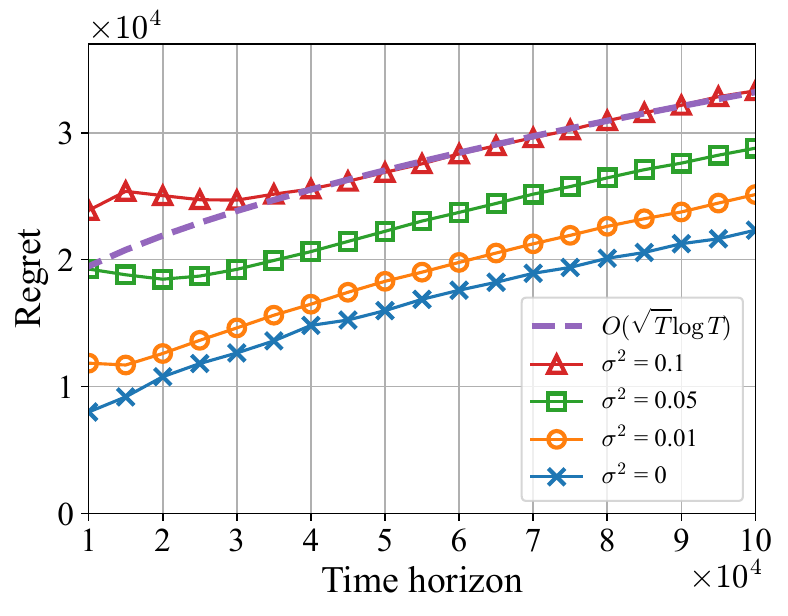}
	\caption{Regret for $\lambda = [1.67, 1.33, 0.33, 1.67]$.}\label{fig:regret-multi-medium}
\end{figure}

\begin{figure}[htbp]
	\centering
	\includegraphics[width=0.9\linewidth]{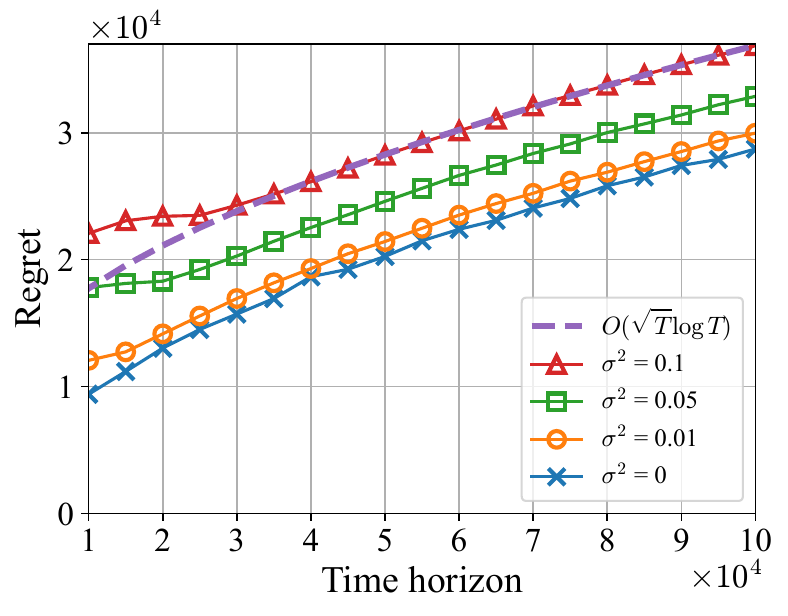}
	\caption{Regret for $\lambda = [2.5, 2.0, 0.5, 2.5]$.}\label{fig:regret-multi-heavy}
\end{figure}

Figures \ref{fig:regret-multi-light}, \ref{fig:regret-multi-medium}, and \ref{fig:regret-multi-heavy} show the resulting regret plotted as a function of time horizon $T=\{10000,20000,...,100000\}$ for different arrival rate vectors $\bs{\lambda}$ corresponding to light, medium, and heavy traffic respectively.
Similar to single user simulations, we show DPOP policy's regrets for different values of noise parameter $\sigma^2 = \{0,0.01, 0.05, 0.1\}$. We also show the $O(\sqrt{T}\log T)$ curve fitted to the DPOP regret of $\sigma^2=0.1$, which demonstrates that the DPOP policy indeed has a sub-linear regret.
From the plots, we can see that the regret increases as the noise level increases, as expected.
Further, similar to single-commodity simulations, we can see that the regret curves of different noise levels are close together when there is heavy traffic (Fig. \ref{fig:regret-multi-heavy}). This is because, when there is heavy traffic, all policies are forced to use all paths irrespective of their cost estimates.
Finally, similar to single-commodity simulations, the oracle policy (the case of $\sigma=0$ in the plots) has a non-zero regret due to the stochastic nature of arrivals.

\subsection{Experiment with $C_B < C_L$}\label{sec:small_CB_experiment}
\begin{figure}[htbp]
	\centering
	\includegraphics[width=0.3\textwidth]{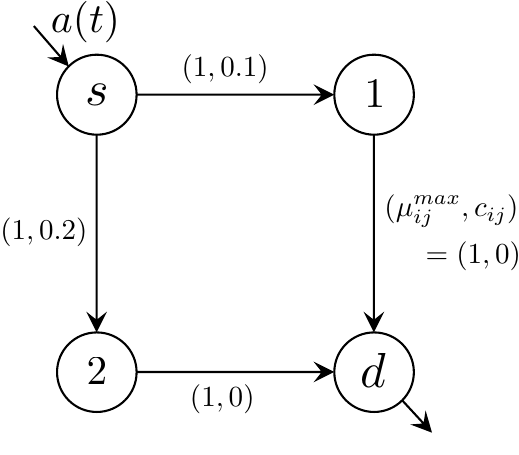}
	\caption{Network topology for $C_B < C_L$ experiment.}\label{fig:toy_topology}
\end{figure}

In Theorem \ref{thm:static_lower_bound} and Corollary \ref{corollary:static}, we obtained a lower bound on the optimal policy's cost when $C_B \geq C_L$. This bound does not hold when $C_B < C_L$ as the best policy can leave packets behind to achieve a lower cost. To study this behavior, we conduct a simulation experiment with $C_B<C_L$. We consider a simple network with 4 nodes and 4 edges shown in Fig. \ref{fig:toy_topology}.
For this network, we have $C_L:=c_{s2}+c_{2d}=0.2$, cost of the most expensive acyclic path. We choose the terminal backlog cost $C_B = 0.15$ such that  $C_B<C_L$. Since Theorem \ref{thm:static_lower_bound}'s bound is not valid anymore, we must derive the exact optimal cost $\pi^* \in \Pi^*$. However, since the optimal policy has complete knowledge of costs and future arrivals, deriving it is difficult. To simplify this, we consider deterministic arrivals $a_s(t) = \lambda = 1.5$ at each time $t$. Therefore, we have $Q_s(t) = \lambda t - \sum_t\tilde{\mu}^{\pi^*}_{s1}(t) - \sum_t\tilde{\mu}^{\pi^*}_{s2}(t)$. Further, since $c_{1d} = 0$, we set ${\mu}^{\pi^*}_{1d}(t) = \mu^{max}_{1d} = 1$ for all $t$. Similarly, since $c_{2d} = 0$, we set ${\mu}^{\pi^*}_{2d}(t) = \mu^{max}_{2d} = 1$ for all $t$. Therefore, since $\mu_{s1}^{\pi^*}(t) \leq \mu_{s1}^{max} = 1$ and $\mu_{s2}^{\pi^*}(t) \leq \mu_{s2}^{max} = 1$, we have $Q_1(t) = Q_2(t) = O(1)$. Combining these, the total cost is
\begin{align*}
	C^{\pi^*}&(T) = \sum_t \tilde{\mu}^{\pi^*}_{s1}(t) c_{s1} + \sum_t \tilde{\mu}^{\pi^*}_{s2}(t) c_{s2}  \\
	&\hspace{1.2cm} + C_{B} \left( \lambda t - \sum_t\tilde{\mu}^{\pi^*}_{s1}(t) - \sum_t\tilde{\mu}^{\pi^*}_{s2}(t) \right) + O(1) \\
	&= \sum_t \tilde{\mu}^{\pi^*}_{s1}(t) \left(c_{s1} - C_B \right) + \sum_t \tilde{\mu}^{\pi^*}_{s2}(t) \left(c_{s2} - C_B \right)  \\
	&\hspace{5.7cm}   + C_{B}  \lambda t + O(1) \\
	&= -0.05 \sum_t \tilde{\mu}^{\pi^*}_{s1}(t) + 0.05 \sum_t \tilde{\mu}^{\pi^*}_{s2}(t) + 0.15 \lambda t + O(1).
\end{align*}

Since $c_{s2} - C_B = 0.05 > 0$ and $c_{s1} - C_B = -0.05 < 0$, the cost minimizer is  $\mu^{\pi^*}_{s2}(t) = 0$ and $\mu^{\pi^*}_{s1}(t) = \mu^{max}_{s1} = 1$. Indeed, since $C_B$ is smaller than the cost $C_L = c_{s2} + c_{2d}$ of the lower path ($s,2,d$), the optimal policy is better off not transmitting any packets on the lower path. This allows the policy to avoid incurring the lower path's transmission costs $C_L = c_{s2} + c_{2d}$, and instead, leave those packets behind in the queues and incur the lower terminal backlog cost $C_B$. As a result, it only transmits on the upper path ($s,1,d$). However, the capacity of the upper path is 1 and since arrival rate is larger than this capacity, $\lambda=1.5>1$, its backlog grows linearly in time, $Q_s(t) = \lambda t - \sum_t\tilde{\mu}^{\pi^*}_{s1}(t) - \sum_t\tilde{\mu}^{\pi^*}_{s2}(t) = 0.5t$. This shows that when $C_B<C_L$, the optimal policy may not be a stable policy.

\begin{figure}[htbp]
	\centering
	\includegraphics[width=0.85\linewidth]{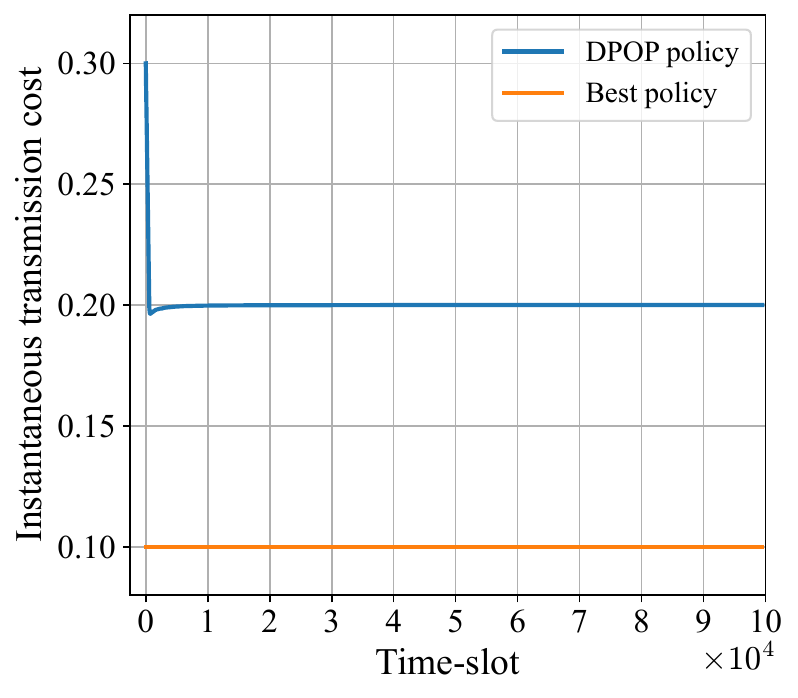}
	\caption{Transmission cost $\sum_{(i,j)\in \mathcal{E}}\Ex[\mu_{ij}^{\pi}(t)]c_{ij}$.}\label{fig:small_CB_cost}
\end{figure}

\begin{figure}[htbp]
	\centering
	\includegraphics[width=0.9\linewidth]{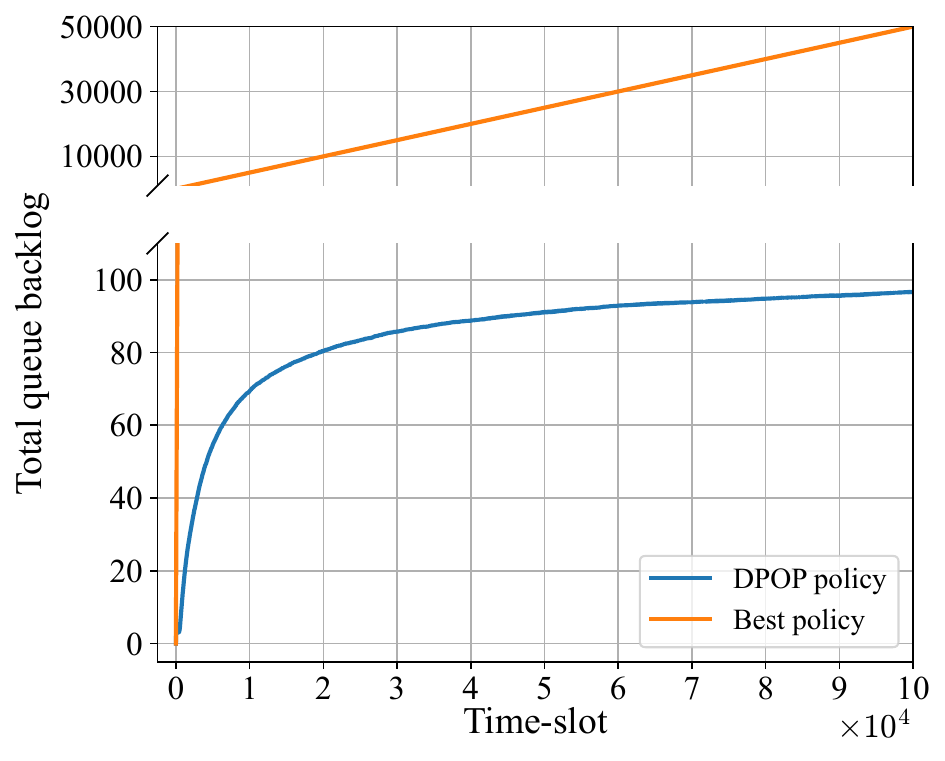}
	\caption{Total queue backlog $\sum_{i\in \mathcal{N}} \Ex[Q_i^{\pi}(t)]$.}\label{fig:small_CB_backlogs}
\end{figure}

We conduct simulations to verify this behavior. We simulate the optimal policy described above for $T=100000$.
Figures \ref{fig:small_CB_cost} and \ref{fig:small_CB_backlogs} show the resulting transmission costs and queue backlogs respectively.
The plots contain the results for both the optimal policy and the DPOP policy. We simulate the DPOP policy with tuning parameters chosen according to Theorem \ref{thm:regret_theorem},  $\beta = 4.5\sigma^2$, $\delta = T^{-{2\sigma^2}/{\beta}}$, and $\nu = \sqrt{T}$. Recall that we obtained this choice of parameters in Theorem \ref{thm:regret_theorem} by optimizing the regret upper bound
 $\sum_t \sum_{(i,j)} \sum_{k} \Ex[\mu^\pi_{ijk}(t) - \mu^{stat}_{ijk} ]c_{ij}
+ C_B\sum_{i} \sum_{k} \Ex \left[Q_{ik}^\pi(T)\right]$
under the assumption that $C_B\geq C_L$. However, this upper bound is no longer valid and as seen from the optimal policy behavior, it is in fact optimal to ignore queue stability. But we designed the DPOP policy's parameters in Theorem \ref{thm:regret_theorem} to stabilize the queues. Therefore, by running the DPOP policy using these parameters, it will still stabilize the network. This can be seen in Fig. \ref{fig:small_CB_backlogs}, where the optimal policy's backlog grows indefinitely and the DPOP policy's backlog is stable. Further, since the optimal policy is minimizing total cost at the expense of queue stability, it is able to achieve a lower transmission cost than DPOP as can be seen in Fig. \ref{fig:small_CB_cost}.

\begin{figure}[htbp]
\centering
\includegraphics[width=0.9\linewidth]{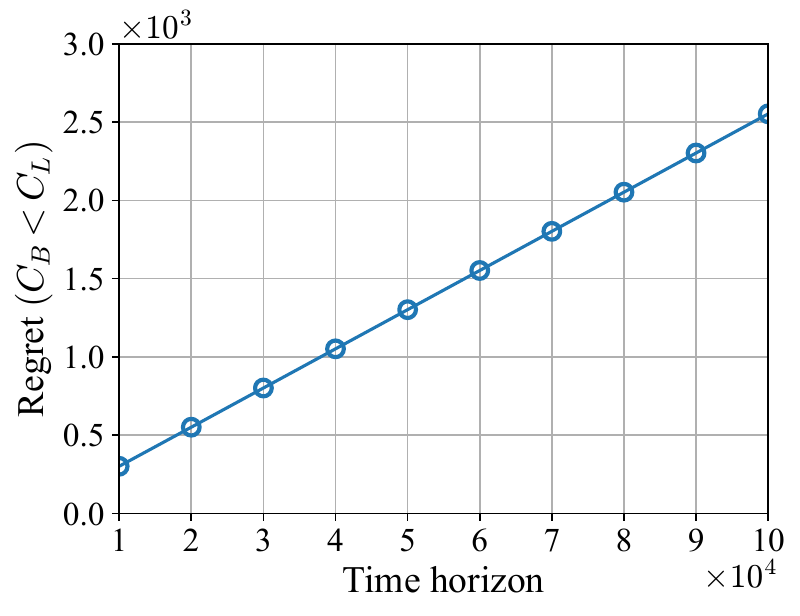}
\caption{DPOP Regret $R^\pi(T)$ when $C_B<C_L$.}\label{fig:small_CB_regret}
\end{figure}

Finally, DPOP's regret $R^\pi(T) = \Ex [C^\pi(T)] - \Ex[ C^{\pi^*}(T)]$ with respect to the optimal policy is shown in Fig. \ref{fig:small_CB_regret}.
Recall that $\Ex [C^\pi(T)]$ is the total unified cost including both transmission costs and terminal backlog cost, $\Ex [C^\pi(T)] = \sum_{t,(i,j),k} \tilde{\mu}^\pi_{ijk}(t)c_{ij} + C_{B}\sum_{i,k} Q_{ik}^\pi(T).$
We can see from the plot that even though DPOP is stabilizing the queues, its regret grows linearly. This is because DPOP is paying higher transmission costs to stabilize the queues, whereas  the optimal policy is paying lower transmission costs at the expense of stability. And since $C_B< C_L$, this results in the optimal policy incurring overall lower unified cost. This example highlights the importance of choosing the correct value of terminal backlog cost $C_B \geq C_L$ to incentivize the policies to stabilize the network.

\section{Conclusion} \label{sec:conclusions}
We considered the problem of making scheduling and routing decisions in multi-commodity queueing networks, where the transmission costs are initially unknown. To design policies that optimize both throughput and cost simultaneously, we defined a novel cost metric in terms of transmission costs and queue backlogs. We showed that any policy's cost is lower bounded by the solution to a static optimization problem. Using this lower-bound, we defined any control policy's regret as the difference between policy's total cost and $T$ times the solution to static optimization problem. Further, using techniques from Lyapunov theory and multi-arm bandits, we designed the Drift Plus Optimistic Penalty Policy.
We showed that this policy achieves a sub-linear regret of order $O(\sqrt{T}\log T)$ and evaluated its performance using simulations. Finally, some potential future research directions include designing control policies when costs are delayed, when costs are a function of queue states, or when congestion control is involved.




{
\appendices

\section{Proof of Lemma \ref{lemma:mean_rates_property} (Effective Rates)}\label{proof:mean_rates_property}

%
%
Recall that $\hat{\mu}^{\pi^*}_{ijk}(t)$ and $\hat{a}_{ik}^{\pi^*}(t)$ are the number of transmitted packets and arrived packets respectively that actually reached the destination by the end of time horizon $T$. Note that these quantities consider only the packets that reached the destination at $T$, and hence do not consider any packets that are still in the queues at $T$. As a result, they adhere to the following flow equation.
$$
\sum_{t=1}^T  \sum_{j:i\in \mc{N}_j} \hat{\mu}^{\pi^*}_{jik}(t) + \sum_{t=1}^T \hat{a}_{ik}^{\pi^*}(t) = \sum_{t=1}^T \sum_{j\in \mc{N}_i} \hat{\mu}^{\pi^*}_{ijk}(t), \; \forall i\neq k.
$$
Taking expectation and dividing by $T$, we obtain (\ref{eq:mean_rates}).
Next, to show (\ref{eq:mean_rates_arrivals}), we denote by $Q^i_{jk}(T)$ the commodity-$k$ packets residing in queue $j$ at time $T$ that originally entered the network at node $i$. Note that $\sum_{i\in \mc{N}} Q^i_{jk}(T)$ is the total number of commodity-$k$ packets at node $j$ at time $T$ i.e. $\sum_{i\in \mc{N}} Q^i_{jk}(T) = Q^{\pi^*}_{jk}(T)$. Now, out of the $a_{ik}(t)$ packets that arrived at node $i$, only $\hat{a}^{\pi^*}_{ik}(t)$ packets actually reached the destination by time $T$, rest of them are distributed across the network. According to our notation, this is represented as
$$
\sum_{t=1}^T a_{ik}(t) = \sum_{t=1}^T \hat{a}^{\pi^*}_{ik}(t) + \sum_{j \in \mc{N}} Q^i_{jk}(T),\quad \forall \; i\neq k.
$$
Summing over $i,k\in\mc{N}$, and taking expectation gives (\ref{eq:mean_rates_arrivals}).
\hfill\IEEEQEDhere

\section{Proof of Theorem \ref{thm:static_lower_bound} (Multi-Commodity Bound)} \label{proof:static_lower_bound_multi}


In this section, we complete the proof of Theorem \ref{thm:static_lower_bound} by extending it to the case of multi-commodity networks. We start by expressing $\mc{P}(\bs{\lambda})$ in its standard form.
\begin{gather*}
	\mc{P}(\bs{\lambda}) = \; \min \bs{c}^T\bs{\mu}, \\
	\text{subject to } M\bs{\mu} = \bs{b}, \text{ and } \bs{\mu} \geq 0
\end{gather*}
where, the decision vector $\boldsymbol{\mu}$ is composed of rates $\mu_{ijk}$'s and slack variables, the objective vector $\bs{c}$ is composed of the costs $c_{ij}$'s and additional zeros corresponding to the slack variables. Further, the constraint vector $\bs{b}$ is composed of arrival rates $\lambda_{ik}$'s and capacities $\mu_{ij}^{max}$'s. Finally, the constraint matrix $M$ is a block matrix composed of the edge adjacency matrix to capture the flow constraints, and identity matrices to capture the capacity constraints. Further, $\mc{P}(\bs{\lambda})$'s dual, denoted by $\mc{D}(\bs{\lambda})$, in standard form is as follows.
\begin{gather*}
	\mc{D}(\bs{\lambda}) := \; \max \bs{p}^T\bs{b}, \\
	\text{subject to }\bs{p}^TM \leq \bs{c}^T.
\end{gather*}

Let $\bs{p^*}$ be an optimal solution to $\mc{D}(\bs{\lambda})$ (at least one optimal solution exists as the problem is bounded). Notice that the feasibility region of $\mc{D}(\bs{\lambda})$ is independent of $\bs{\lambda}$, hence $\bs{p^*}$ is a feasible solution to $\mc{D}(\bar{\bs{\lambda}})$. Using weak duality at $\bar{\bs{\lambda}}$, we have $\mc{P}(\bar{\bs{\lambda}}) \geq \bs{p^*}^T\bar{\bs{b}}$ where, $\bar{\bs{b}}$ is the new constraint vector composed of effective arrival rates $\bar{\lambda}_{ik}^{\pi^*}$'s  and capacities $\mu_{ij}^{max}$'s. Here, if there are multiple optimal solutions $\bs{p^*}$'s, we pick the one that maximizes $\bs{p^*}^T\bs{\bar{b}}$ so that the inequality is tight. Further, using strong duality at $\bs{\lambda}$, we have $\mc{P}(\bs{\lambda}) = \mc{D}(\bs{\lambda}) = \bs{p^*}^T{\bs{b}} $. Combining these two results, $\mc{P}(\bar{\bs{\lambda}}) - \mc{P}(\bs{\lambda}) \geq \bs{p^*}^T\left(\bar{\bs{b}} - \bs{b}\right)$. Finally, since $\bar{\bs{b}}$ differs from $\bs{b}$ only at indices of $\lambda_{ik}$'s (and remains the same at the indices of $\mu_{ij}^{max}$'s), with a slight abuse of notation, we rewrite this result as $\mc{P}(\bar{\bs{\lambda}}) - \mc{P}(\bs{\lambda}) \geq \bs{p^*}^T\left(\bar{\bs{\lambda}} - \bs{\lambda}\right).$ 
Plugging this in (\ref{eq:cost_inequlality}), we have
\begin{align}\label{eq:cost_inequality_dual}
	\Ex[C^{\pi^*}(T)] - T \mc{P}(\bs{\lambda}) \geq T \big( C_B\bs{1} -  \bs{p^*} \big)^T \big(\bs{\lambda} - \bar{\bs{\lambda}} \big)
\end{align}
where, $\bs{1}$ is a vector of all ones. Further, since the problem is in standard form, the optimal solution is of the form $\bs{p^*}^T = \bs{c}_B^TM_B^{-1}$, where $B$ is a set of basis columns. Now, pick $C_L := \| \max_B \bs{c}_B^TM_B^{-1} \|_\infty$, where maximization is done over all possible basis columns of the constraint matrix $M$, and $\|\cdot\|_\infty$ represents the infinity norm. Notice that $C_L$ is finite and dependent only on costs and network topology. For this choice of $C_L$, we have $C_L\bs{1} \geq \bs{p^*}$.
Finally, plugging this in (\ref{eq:cost_inequality_dual}), using $C_B \geq C_L$, and using the fact that $\bs{\lambda} \geq \bar{\bs{\lambda}}$, we have $\Ex[C^{\pi^*}(T)] - T \sum_{(i,j)\in \mc{E}} \sum_{k\in\mc{N}} \mu^{stat}_{ijk}c_{ij} \geq 0$.
\hfill\IEEEQEDhere

\section{Proof of Lemma \ref{lemma:regret_decomp} (Regret Decomposition)}\label{proof:regret_decomp}

%
From Corollary \ref{corollary:static}, we can bound the regret as
\begin{align*}
		R^{\pi_D}(T) \leq \sum_{t=1}^T \sum_{(i,j)\in \mc{E}} \sum_{k\in\mc{N}} \Ex & \left[ \left(\mu^{\pi_D}_{ijk}(t) - \mu^{stat}_{ijk} \right) c_{ij}\right] \\
		& + C_B\sum_{i\in \mc{N}} \sum_{k\in\mc{N}} \Ex \left[Q_{ik}^{\pi_D}(T)\right].
\end{align*}
Now, we split the term inside first summation by conditioning it on the event $A$ and its compliment $\bar{A}$.
\begin{align}
	\EX \Big[ \Big(\mu^{\pi_D}_{ijk}(t) - & \mu^{stat}_{ijk} \Big) c_{ij} \Big] = \EX \left[ \left(\mu^{\pi_D}_{ijk}(t) - \mu^{stat}_{ijk} \right)c_{ij} \;|\; A \right] \PR[{A}] \nonumber \\
	&\hspace{4mm} + \EX \left[ \left(\mu^{\pi_D}_{ijk}(t) - \mu^{stat}_{ijk} \right)c_{ij} \;|\; \bar{A} \right] \PR[\bar{A}]. \label{eq:regret_decomp_1}
\end{align}

We first bound the expectation conditioned on $A$  by adding and subtracting the optimistic estimates $ \hat{c}_{ij}(t)$'s to $c_{ij}$'s,
\begin{align*}
	\EX \Big[ \Big(\mu^{\pi_D}_{ijk}(t) - \mu^{stat}_{ijk} \Big)c_{ij} \;|\; A \Big] = \EX\left[(\mu^{\pi_D}_{ijk}(t) - \mu^{stat}_{ijk}) \hat{c}_{ij}(t) \;|\; A \right] \\
	  + \EX\left[(\mu^{\pi_D}_{ijk}(t) - \mu^{stat}_{ijk})(c_{ij}- \hat{c}_{ij}(t)) \;|\; A \right]
\end{align*}
Under event $A$, we have $c_{ij} \geq   \bar{c}_{ij}(t) - \sqrt{{\beta\log (t/\delta)}/{N_{ij}(t)}} = \hat{c}_{ij}(t)$. Therefore, we have $\EX[\mu^{stat}_{ijk}(c_{ij}- \hat{c}_{ij}(t)) | A ] \geq 0$, and
\begin{align*}
	\EX \Big[ \Big(\mu^{\pi_D}_{ijk}(t) - \mu^{stat}_{ijk} \Big)c_{ij} \;|\; A \Big] \leq \EX\left[(\mu^{\pi_D}_{ijk}(t) - \mu^{stat}_{ijk}) \hat{c}_{ij}(t) \;|\; A \right] \\
	+  \EX\left[\mu^{\pi_D}_{ijk}(t)(c_{ij}- \hat{c}_{ij}(t)) \;|\; A \right]
\end{align*}

Next, we bound expectation conditioned on $\bar{A}$. Notice that since $\mu^{\pi_D}_{ijk}(t) - \mu^{stat}_{ijk} \leq  \mu^{max}_{ij}$ and since $c_{ij}\in[0,C_{max}]$,
$
\EX \left[ \left(\mu^{\pi_D}_{ijk}(t) - \mu^{stat}_{ijk} \right)c_{ij} \;|\; \bar{A} \right] \leq C_{max} \mu^{max}_{ij}.
$
Plugging these back in (\ref{eq:regret_decomp_1}) and using $\PR[A]\leq 1$,
\begin{align*}
	\EX \Big[ \Big(\mu^{\pi_D}_{ijk}(t)& - \mu^{stat}_{ijk} \Big) c_{ij} \Big]  \leq \EX\left[(\mu^{\pi_D}_{ijk}(t) - \mu^{stat}_{ijk}) \hat{c}_{ij}(t) \;|\; A \right] \Pb[A] \\
	&+ 	\EX\left[\mu^{\pi_D}_{ijk}(t)(c_{ij}- \hat{c}_{ij}(t)) \;|\; A \right] + C_{max} \mu^{max}_{ij} \PR \left[\bar{A} \right].
\end{align*}
Finally, plugging this in $R^{\pi_D}(T)$ completes the proof.
\hfill\IEEEQEDhere

\section{Proof of Proposition \ref{prop:exploration_regret}}\label{proof:exploration_regret}


%

Recall $R^{\pi_D}_1(T) := \sum_{t=1}^T\sum_{(i,j)\in \mathcal{E}} \sum_{k \in \mc{N}} \Ex [ \mu^{\pi_D}_{ijk}(t)(c_{ij} - \hat{c}_{ij}(t)) \;|\; A ]$. Adding and subtracting $ \bar{c}_{ij}(t)$'s to $c_{ij}$'s, we obtain
\begin{multline*}
	R^{\pi_D}_1(T) \leq \sum_{t=1}^T\sum_{(i,j)\in \mathcal{E}}\sum_{k \in \mc{N}} \EX \Big[ \mu^{\pi_D}_{ijk}(t) \Big(|c_{ij} -  \bar{c}_{ij}(t)| \\
	+ | \bar{c}_{ij}(t) - \hat{c}_{ij}(t)| \Big) \;\Big|\; A \Big].
\end{multline*}
Under the event $A$, we have by $A$'s definition,
$
\left| c_{ij} -  \bar{c}_{ij}(t) \right| \leq \sqrt{\frac{\beta\log (t/\delta)}{N_{ij}(t)}} \leq \sqrt{\frac{\beta\log (T/\delta)}{N_{ij}(t)}}.
$
Further, by $\hat{c}_{ij}(t)$'s definition, we have
$
\left| \hat{c}_{ij}(t) -  \bar{c}_{ij}(t) \right| = \sqrt{\frac{\beta\log (t/\delta)}{N_{ij}(t)}} \leq \sqrt{\frac{\beta\log (T/\delta)}{N_{ij}(t)}}.
$
Plugging these inequalities in $R^{\pi_D}_1(T)$'s bound,
\begin{equation*}
	R^{\pi_D}_1(T) \leq 2\sum_{t=1}^T\sum_{(i,j)\in \mathcal{E}}\sum_{k\in\mc{N}} \EX \biggl[ \mu^{\pi_D}_{ijk}(t)\sqrt{\frac{\beta\log (T/\delta)}{N_{ij}(t)}} \biggm| A \biggr].
\end{equation*}
\begin{lemma}\label{lemma:bound_on_x}
	$\sum_{t}\sum_{ij} \sum_{k}  {\mu_{ijk}^{\pi_D}(t)}/{\sqrt{N_{ij}(t)}} = O(\sqrt{T\log T}).$
\end{lemma}

Plugging Lemma \ref{lemma:bound_on_x}'s result in $R^{\pi_D}_1(T)$'s bound gives us the desired result. We are now left with proving Lemma  \ref{lemma:bound_on_x}.

\paragraph*{Proof of Lemma \ref{lemma:bound_on_x}}
Let $x_{ijk}(t) := \mu_{ijk}^{\pi_{D}}(t)/\mu_{ij}^{max}$. Expressing in terms of $x_{ijk}(t)$ and using Cauchy--Schwarz,
\begin{align}
	\sum_{(i,j)\in \mathcal{E}}\sum_{k \in \mc{N}} &\sum_{t=1}^T \frac{\mu_{ijk}^{\pi_D}(t)}{\sqrt{N_{ij}(t)}} =\sum_{(i,j)\in \mathcal{E}} \sum_{k \in \mc{N}} \sum_{t=1}^T \frac{\mu_{ij}^{max}x_{ijk}(t)}{\sqrt{N_{ij}(t)}} \nonumber \\
	&\leq \sum_{(i,j)\in \mathcal{E}} \sum_{k \in \mc{N}} \sqrt{\sum_{t=1}^T(\mu_{ij}^{max})^2 \sum_{t=1}^T \frac{x_{ijk}^2(t)}{N_{ij}(t)}}. \label{eq:mu_by_N}
\end{align}
Further, by definition, $x_{ijk}(t) \in [0,1]$ and $N_{ij}(t) \geq 1$, hence ${x_{ijk}^2(t)} / {N_{ij}(t)} \in [0,1]$. Using the formula  $\forall y\in [0,1], \;y\leq 2\log(1+y)$, we can see that
$
\frac{x_{ijk}^2(t)}{N_{ij}(t)} \leq 2\log \left(1+\frac{x_{ijk}^2(t)}{{N_{ij}(t)}}\right).
$
Now, recall $N_{ij}(t)$ 's update equation, $N_{ij}(t+1) = N_{ij}(t) + e_{ij}(t)$. The indicator variable $e_{ij}(t) = 0$ only if $\mu_{ijk}^{\pi_{D}}(t) = 0$. Hence, we have $e_{ij}(t) \geq x^2_{ijk}(t)$. Using this,
\begin{align*}
	N_{ij}(t+1) &= N_{ij}(t) + e_{ij}(t) \geq N_{ij}(t) + x^2_{ijk}(t) \\
	&= N_{ij}(t)(1 + {x^2_{ijk}(t)}/{N_{ij}(t)}).
\end{align*}
Taking $\log(\cdot)$ and telescoping over time $t=1,2,...,T-1$,
$
	\log N_{ij}(T+1) \geq \log N_{ij}(1) + \sum_{t=1}^T\log \left(1 + \frac{x^2_{ijk}(t)}{N_{ij}(t)} \right).
$
Hence, using the fact that $N_{ij}(1) = 1$ and $N_{ij}(T+1) \leq T+1$,
$$\sum_{t=1}^T \frac{x_{ijk}^2(t)}{N_{ij}(t)} \leq 2\sum_{t=1}^T\log \left(1 + \frac{x^2_{ijk}(t)}{N_{ij}(t)} \right) = O(\log T).$$
Finally, notice that $ \sum_{t=1}^T(\mu_{ij}^{max})^2 = O(T)$. Plugging these results back in (\ref{eq:mu_by_N}) completes the proof.
\hfill\IEEEQEDhere

\section{Proof of Proposition \ref{prop:regret_3_bound}}\label{proof:regret_3_bound}
%
Recall the definition of this regret component
$R^{\pi_D}_2(T) := T|\mc{N}| \sum_{(i,j)\in \mathcal{E}}\mu_{ij}^{max}C_{max} \PR[\bar{A}].$
We analyze the probability term $\PR(\bar{A})$. Using the union bound,
\begin{align}
	\PR(&\bar{A}) = \PR \biggl[\exists t, \exists (i,j) : |c_{ij} -  \bar{c}_{ij}(t)| > \sqrt{\frac{\beta\log (t/\delta)}{N_{ij}(t)}} \biggr] \nonumber \\
		&\leq \sum_{t=1}^T \sum_{(i,j)\in \mathcal{E}} \PR\biggl[ |c_{ij} -  \bar{c}_{ij}(t)| > \sqrt{\frac{\beta\log (t/\delta)}{N_{ij}(t)}} \biggr]. \label{eq:probab_error}
\end{align}

Define $\mathcal{T}_{ij}(t) = \{ \tau \in\{0, 1, ..., t-1\}: e_{ij}(\tau) = 1 \}$ as the set of time-slots until $t$ in which a feedback value was received. It follows that $|\mathcal{T}_{ij}(t)| = N_{ij}(t)$. Moreover, the noisy feedback values received until time $t$ for each edge $(i,j)$ are $\forall \tau\in\mathcal{T}_{ij}(t), \; \widetilde{c}_{ij} (\tau) = c_{ij} + \eta_{ij}(\tau)$. Hence, for $t\geq 1$, the average cost estimate $\bar{c}_{ij}(t)$ is given by
$
 \bar{c}_{ij}(t) =  \frac{\sum_{\tau\in\mathcal{T}_{ij}(t)} \widetilde{c}_{ij}(\tau)}{|\mathcal{T}_{ij}(t)|} =c_{ij} + \frac{\sum_{\tau\in\mathcal{T}_{ij}(t)} \eta_{ij}(\tau)}{N_{ij}(t)}.
$
where, the second equality is valid since $N_{ij}(t)\geq1$, $\forall(i,j)$ and $t\geq 1$. Denoting $\theta_t:=\sqrt{{\beta\log (t/\delta)}/{N_{ij}(t)}}$, we now have
\begin{align*}
	\PR \left[|c_{ij} -  \bar{c}_{ij}(t)| > \theta_t \right] = \PR \left[  \frac{|\sum_{\tau\in\mathcal{T}_{ij}(t)} \eta_{ij}(\tau)|}{N_{ij}(t)} > \sqrt{\frac{\beta\log \frac{t}{\delta} }{N_{ij}(t)}} \right] \\
		\leq \sum_{n=1}^{t-1} \PR \left[   \Bigg|\sum_{\tau\in\mathcal{T}_{ij}(t)} \eta_{ij}(\tau) \Bigg| > \sqrt{ n\beta\log (t/\delta) } \;\Bigg|\; N_{ij}(t) = n \right]
\end{align*}
where, we get the inequality by conditioning on number of observations $N_{ij}(t) = |\mathcal{T}_{ij}(t)| = n$. Now, since $\eta_{ij}(t)$ are independent $\sigma$-sub-Gaussian, Hoeffding's bound \cite{wainwright} gives us
\begin{align*}
	\PR \left[   \Bigg|\sum_{\tau\in\mathcal{T}_{ij}(t)} \eta_{ij}(\tau) \Bigg| > \sqrt{ n\beta\log (t/\delta) } \;\Bigg|\; |\mathcal{T}_{ij}(t)| = n \right] \\
	\leq 2 e^{-\frac{ n\beta\log (t/\delta)}{2n\sigma^2}} = 2e^{-\frac{\beta}{2\sigma^2}\log ({t}/{\delta})}.
\end{align*}

Plugging this in previous inequality with $\delta = T^{-2\sigma^2/\beta}$,
$$
	\PR\left[|c_{ij} -  \bar{c}_{ij}(t)| > \theta_t \right] \leq \sum_{n=1}^{t-1} 2e^{-\frac{\beta}{2\sigma^2}\log ({t}/{\delta})} =  \frac{2t^{1-\beta/2\sigma^2}}{T}.
$$
Plugging this in (\ref{eq:probab_error}), we have $\PR[\bar{A}] \leq \frac{2|\mc{E}|}{T} \sum_{t=1}^T t^{1-\beta/2\sigma^2}.$ Taking $\beta > 4\sigma^2$, we have $\sum_{t=1}^T t^{1-\beta/2\sigma^2} = O(1)$. Hence,
$
	\PR[\bar{A}] \leq \frac{2|\mc{E}|}{T} \sum_{t=1}^T t^{1-\beta/2\sigma^2} = O(1/T).
$
Finally, $R^{\pi_D}_2(T) = T|\mc{N}|\sum_{ij\in \mathcal{E}}\mu_{ij}^{max}C_{max} \PR[\bar{A}] = O(1).$
\hfill\IEEEQEDhere

\section{Proof of Proposition \ref{prop:cost_gap_to_stat}}\label{proof:cost_gap_to_stat}
%
We first bound the conditional drift-plus-optimistic-penalty.\hspace*{-2pt}
\begin{lemma}\label{lemma:drift_bound}	Given $\bs{Q}^{\pi_D}(t)$ and $\{ \hat{c}_{ij}(t): {(i,j)\in\mc{E}}\}$, policy $\pi_D$'s conditional drift-plus-optimistic-penalty is bounded as
	\begin{multline}\label{eq:dpp_bound_lemma}
		  \Ex_{|Q, \hat{c}} \biggl[ \Delta\Phi^{\pi_D}(t) + \nu \sum_{(i,j)\in \mc{E}} \sum_{k \in \mc{N}} \mu_{ijk}^{\pi_D}(t) \hat{c}_{ij}(t) \biggr] \\
		  \leq B + \nu \sum_{k \in \mc{N}} \sum_{(i,j)\in \mc{E}}\mu_{ijk}^{stat} \hat{c}_{ij}(t).
	\end{multline}
    where, $\Ex_{| Q, \hat{c}}[\cdot] := \Ex[\cdot \;|\; \bs{Q}^{\pi_D}(t), \{ \hat{c}_{ij}(t):{(i,j)\in\mc{E}}\}]$, and $B$ is a constant independent of $T$.
\end{lemma}

Taking $\Ex[\cdot|A]$ on both sides of \eqref{eq:dpp_bound_lemma} and using total law of expectations,
$
	\Ex \left[ \Delta\Phi^{\pi_D}(t) + \nu \sum_{(i,j),k} \mu_{ijk}^{\pi_D}(t) \hat{c}_{ij}(t) \mid A \right]
	\leq B + \nu \sum_{k,(i,j)} \Ex \left[ \mu_{ijk}^{stat} \hat{c}_{ij}(t) \mid A \right].
$
Summing over $t$, we get
\begin{align}
	R^{\pi_D}_3(T&) = \sum_{t=1}^T \sum_{k \in \mc{N}} \sum_{(i,j)\in \mc{E}} \EX [(\mu^{\pi_D}_{ijk}(t) - \mu_{ijk}^{stat}) \hat{c}_{ij}(t) \mid A] \Pb[A] \nonumber \\
	&\leq \left(  \frac{BT - \Ex[ \Phi^{\pi_D}(T+1) \;|\; A]}{\nu}  \right)\Pb[A] \leq \frac{BT}{\nu}. \label{eq:R_3_dep_on_CB}
\end{align}
where, the last inequality is because $ \Phi^{\pi_D}(t)\geq 0, \forall t$, and $\Pb[A]\leq 1$. Plugging in $\nu = \sqrt{T}$, we get the desired result. We are only left with proving Lemma \ref{lemma:drift_bound} to complete the proof.

\begin{IEEEproof}[Proof of Lemma \ref{lemma:drift_bound}]
Recall from Section \ref{sec:algorithm} that policy $\pi_D$ minimizes the bound on drift-plus-optimistic-penalty $\hat{L}^{\pi_D}(t)$. Therefore, by comparing $\pi_D$'s drift-plus-optimistic-penalty $\hat{L}^{\pi_D}(t)$ against $\mu_{ijk}^{stat}$, the solution to $\mc{P}$, we get
\begin{align*}
   	\Ex& \Biggl[ \Delta\Phi^{\pi_D}(t) + \nu \sum_{(i,j)\in \mc{E}} \sum_{k \in \mc{N}} \mu_{ijk}^{\pi_D}(t) \hat{c}_{ij}(t) \mid \bs{Q}^{\pi_D}(t), \{ \hat{c}_{ij}(t)\}\Biggr] \\
   	 &\leq B + \sum_{k\in\mc{N}} \sum_{i\in\mc{N}}  \lambda_{ik} Q_{ik}^{\pi_D}(t) \\
   	&\hspace{0.75cm} + \sum_{k\in\mc{N}} \sum_{(i,j)\in \mc{E}} \mu_{ijk}^{\pi_D}(t)\left(Q^{\pi_D}_{jk}(t) - Q_{ik}^{\pi_D}(t) + \nu \hat{c}_{ij}(t)  \right) \\
   	&\leq B + \sum_{k\in\mc{N}} \sum_{i\in\mc{N}}  \lambda_{ik} Q_{ik}^{\pi_D}(t) \\
   	&\hspace{0.75cm} + \sum_{k\in\mc{N}} \sum_{(i,j)\in \mc{E}} \mu_{ijk}^{stat}\left(Q^{\pi_D}_{jk}(t) - Q_{ik}^{\pi_D}(t) + \nu \hat{c}_{ij}(t)  \right) \\
   	&= B + \nu \sum_{k \in \mc{N}} \sum_{(i,j)\in \mc{E}}\mu_{ijk}^{stat} \hat{c}_{ij}(t) \\
   	&\hspace{0.75cm}+ \sum_{k \in \mc{N}} \sum_{i\in \mc{N}}Q_{ik}^{\pi_D}(t) \Biggl[ \lambda_{ik} + \sum_{j:i\in \mc{N}_j} \mu_{jik}^{stat} - \sum_{j\in \mc{N}_i} \mu_{ijk}^{stat}  \Biggr].
\end{align*}
The first inequality is obtained by using the queue evolution expression, the second inequality is due to DPOP policy being the minimizer, and finally the equality is obtained by rearranging the summation. From the constraints of $\mc{P}$, we have $\lambda_{ik} + \sum_{j:i\in \mc{N}_j} \mu_{jik} \leq \sum_{j\in \mc{N}_i} \mu_{ijk}, \; \forall i\neq k$. Using this in the above inequality concludes the proof of lemma.
\end{IEEEproof}

\section{Proof of Proposition \ref{prop:queueing_regret}}\label{proof:queueing_regret}
%
To prove this result, we employ the \emph{drift lemma} from \cite{neely_convex_opt}.\hspace{-2pt}
\begin{lemma}[Drift Lemma \cite{neely_convex_opt}]\label{lemma:neely_lemma}
Let $\{Z(t) : t\geq0\}$ be a discrete time stochastic process adapted to a filtration $\{\mc{F}(t) : t\geq0\}$ with $Z(t)=0$ and $\mc{F}(0) = \{\phi,\Omega\}$. If there exists an integer $t_0>0$, real constants $\theta>0$, $\delta_{max}>0$, and $0<\zeta<\delta_{max}$ such that the following holds $\forall t\geq1$,
\begin{enumerate}
	\item $|Z(t+1)-Z(t)|\leq \delta_{max}$,
	\item $\EX[Z(t+t_0)-Z(t)|\mc{F}(t)]\leq t_0\delta_{max}$, if $Z(t)<\theta$,
	\item $\EX[Z(t+t_0)-Z(t)|\mc{F}(t)]\leq -t_0\zeta$, if $Z(t)\geq\theta$.
\end{enumerate}
Then, $\EX[Z(t)] \leq \theta + t_0\delta_{max} + t_0 \frac{4\delta^2_{max}}{\zeta}\log\frac{8\delta^2_{max}}{\zeta^2}$, $\forall t\geq1$.

\end{lemma}

We use this lemma taking $Z(t) = \sqrt{2\Phi^{\pi_D}(t)}$ and $t_0 = 1$. Using the boundedness of arrivals and transmissions, it can be verified that $Z(t)$ satisfies the first two conditions of the lemma i.e. there exists a constant $\delta_{max} > 0$ such that $|Z(t+1)-Z(t)|\leq \delta_{max}$. We now verify that $Z(t)$ satisfies the final condition. Recall $\exists\;\epsilon>0$ such that $\boldsymbol{\lambda} + \epsilon\mathbf{1} \in \Lambda(G)$. Hence, $\exists\; \bs{\mu}^\epsilon := \{\mu_{ijk}^{\epsilon} : (i,j)\in\mc{E}, k\in\mc{N} \}$ such that
\begin{equation}\label{eq:epsilon_mu}
	\begin{Bmatrix}
		\sum_{j:i\in \mc{N}_j} \mu_{jik}^\epsilon + \lambda_{ik} +\epsilon \leq \sum_{j\in \mc{N}_i} \mu_{ijk}^\epsilon, \; \forall i\neq k, \\ 
		\sum_{k \in \mc{N}} \mu_{ijk}^\epsilon \leq \mu_{ij}^{max}, \; \forall (i,j)\in \mc{E}, \\
		\mu_{ijk}^\epsilon \geq 0, \; \forall (i,j) \in \mc{E}, k\in\mc{N}.
	\end{Bmatrix}
\end{equation}
Since $\pi_D$ minimizes the bound on drift-plus-optimistic-penalty $\hat{L}^{\pi_D}(t)$, we compare ${\pi_D}$'s drift-plus-optimistic-penalty bound with $\mu_{ijk}^{\epsilon}$, similar to Lemma \ref{lemma:drift_bound}'s proof.
\begin{align}
	\Ex\Biggl[& \Delta\Phi^{\pi_D}(t) + \nu \sum_{(i,j)\in \mc{E}} \sum_{k \in \mc{N}} \mu_{ijk}^{\pi_D}(t) \hat{c}_{ij}(t) \mid \bs{Q}^{\pi_D}(t), \{ \hat{c}_{ij}(t) \} \Biggr] \nonumber \\
	&\leq B + \sum_{k\in\mc{N}} \sum_{i\in\mc{N}}  \lambda_{ik} Q_{ik}^{\pi_D}(t) \nonumber \\
	&\hspace{0.75cm} + \sum_{k\in\mc{N}} \sum_{(i,j)\in \mc{E}} \mu_{ijk}^{\epsilon}\left(Q^{\pi_D}_{jk}(t) - Q_{ik}^{\pi_D}(t) + \nu \hat{c}_{ij}(t)  \right) \nonumber \\
	&\leq B + \nu \sum_{k \in \mc{N}} \sum_{(i,j)\in \mc{E}}\mu_{ijk}^{\epsilon} \hat{c}_{ij}(t) - \epsilon \sum_{k \in \mc{N}} \sum_{i\in \mc{N}}Q_{ik}^{\pi_D}(t) \label{eq:epsilon_bound}
\end{align}
The first inequality is obtained using the queue evolution equations, the second inequality is due to DPOP policy being the minimizer of drift-plus-optimistic penalty, and the final inequality is obtained by rearranging the summation and using (\ref{eq:epsilon_mu}).
Further, since transmissions and costs are bounded,
$
\mu_{ijk}^{\epsilon} \hat{c}_{ij}(t) \leq \mu_{ij}^{max}C_{max},  \forall (i,j), k.
$
Since $ \hat{c}_{ij}(t) =  \bar{c}_{ij}(t) - \sqrt{ \frac {\beta\log (t/\delta)}{N_{ij}(t)}} \geq - \sqrt{\beta\log({T}/{\delta})}$, we have
$
\EX[\mu_{ijk}^{\pi_D}(t) \hat{c}_{ij}(t)]  \geq -\mu_{ij}^{max}\sqrt{ \beta \log({T}/{\delta)}},  \forall (i,j), k.
$
Using these cost bounds in  (\ref{eq:epsilon_bound}) and rearranging, we have
\begin{align}\label{eq:drift_epsilon_inequality}
	\EX \left[ \Delta\Phi^{\pi_D}(t) \mid \bs{Q}^{\pi_D}(t) \right] &\leq \epsilon\theta/2 - \epsilon\sum_{i\in \mc{N}} \sum_{k \in \mc{N}} Q_{ik}^{\pi_D}(t)
\end{align}
where, we have collected all the terms that do not depend on $\bs{Q}^{\pi_D}(t)$ into a positive constant $\theta > 0$ defined as
\begin{equation}
\theta := \frac{2B}{\epsilon}+\frac{2\nu|\mc{N}|}{\epsilon} \hspace{-1.5mm} \sum_{(i,j)\in \mc{E}}\mu_{ij}^{max} \hspace{-0.5mm} \left(C_{max} \hspace{-0.5mm} + \hspace{-0.5mm} \sqrt{ \beta \log({T}/{\delta}) }\right).\label{eq:theta_defn}
\end{equation}

Using the $L^1$--$L^2$ norm inequality, we have
$
\sum_{i,k} Q_{ik}^{\pi_D}(t) \geq \sqrt{\sum_{i,k} Q_{ik}^{\pi_D}(t)^2} = \sqrt{2\Phi^{\pi_D}(t)} = Z(t).
$
Plugging this in (\ref{eq:drift_epsilon_inequality}), and expressing $\Delta\Phi^{\pi_D}(T)$ in terms of $Z(t) = \sqrt{2\Phi^{\pi_D}(t)}$, we obtain
$
\EX \left[ Z^2(t+1) -  Z^2(t) \mid \bs{Q}^{\pi_D}(t)  \right] \leq \epsilon\theta  - 2\epsilon Z(t).
$
As we are working with upper bounds, we can make a perfect square by adding an $\epsilon^2/4$ term as
\begin{align*}
	\EX \big[ Z^2(t+1) \mid \bs{Q}^{\pi_D} & (t) \big] \leq \epsilon\theta  - \epsilon Z(t) - \epsilon Z(t) + Z^2(t) \\
	&\leq \epsilon\left(\theta-Z(t)\right) + \epsilon^2/4 - \epsilon Z(t) + Z^2(t). \\
	& = \epsilon\left(\theta-Z(t)\right) + \left(Z(t)- \epsilon/2\right)^2.
\end{align*}

Further, using Jensen's inequality, if $Z(t)\geq \theta$, we have,
$
	\left( \EX \left[ Z(t+1) \mid \bs{Q}^{\pi_D}(t) \right] \right)^2  \leq \EX \left[ Z^2(t+1) \mid \bs{Q}^{\pi_D}(t) \right] \leq \left(Z(t)- \epsilon/2\right)^2.
$
Hence, we have $\EX [ Z(t+1)-Z(t) \; | \; \bs{Q}^{\pi_D}(t) ] \leq -\epsilon/2$ for small values of  $\epsilon$.
As a result, we have shown that $Z(t) = \sqrt{2\Phi^{\pi_D}(t)}$ meets all the conditions of  Lemma \ref{lemma:neely_lemma}. Hence, employing Lemma \ref{lemma:neely_lemma} with $\zeta = \epsilon/2$, $t_0=1$, $\theta$ defined in (\ref{eq:theta_defn}), and ignoring the constants, we obtain
$
	\EX[\sqrt{2\Phi^{\pi_D}(T)}] = \EX[Z(T)] = O(\theta) = O(\nu\sqrt{\log(T/\delta)}).
$
Plugging in parameter $\delta = T^{-{2\sigma^2}/{\beta}}$, we have $\EX [\sqrt{\Phi^{\pi_D}(T)}] = O(\nu \sqrt{ \log T})$.
Finally, we can get the required bound on the regret component using Cauchy--Schwarz inequality, and setting $\nu = \sqrt{T}$ as
\begin{align}
	R&^{\pi_D}_4(T) = C_B \EX \left[ \sum_{i,k} Q^{\pi_D}_{ik}(T) \right] \leq C_B \sqrt{2} |\mc{N}| \EX \left[ \sqrt{\Phi^{\pi_D}(T)} \right] \nonumber\\
	&= C_B O(\nu \sqrt{ \log T})  = O(\sqrt{T \log T}). \hspace*{2cm} \IEEEQEDhere \label{eq:R_4_dep_on_CB}
\end{align}

%
%

}

\begin{IEEEbiography}[{\includegraphics[width=1in,height=1.25in,clip,keepaspectratio]{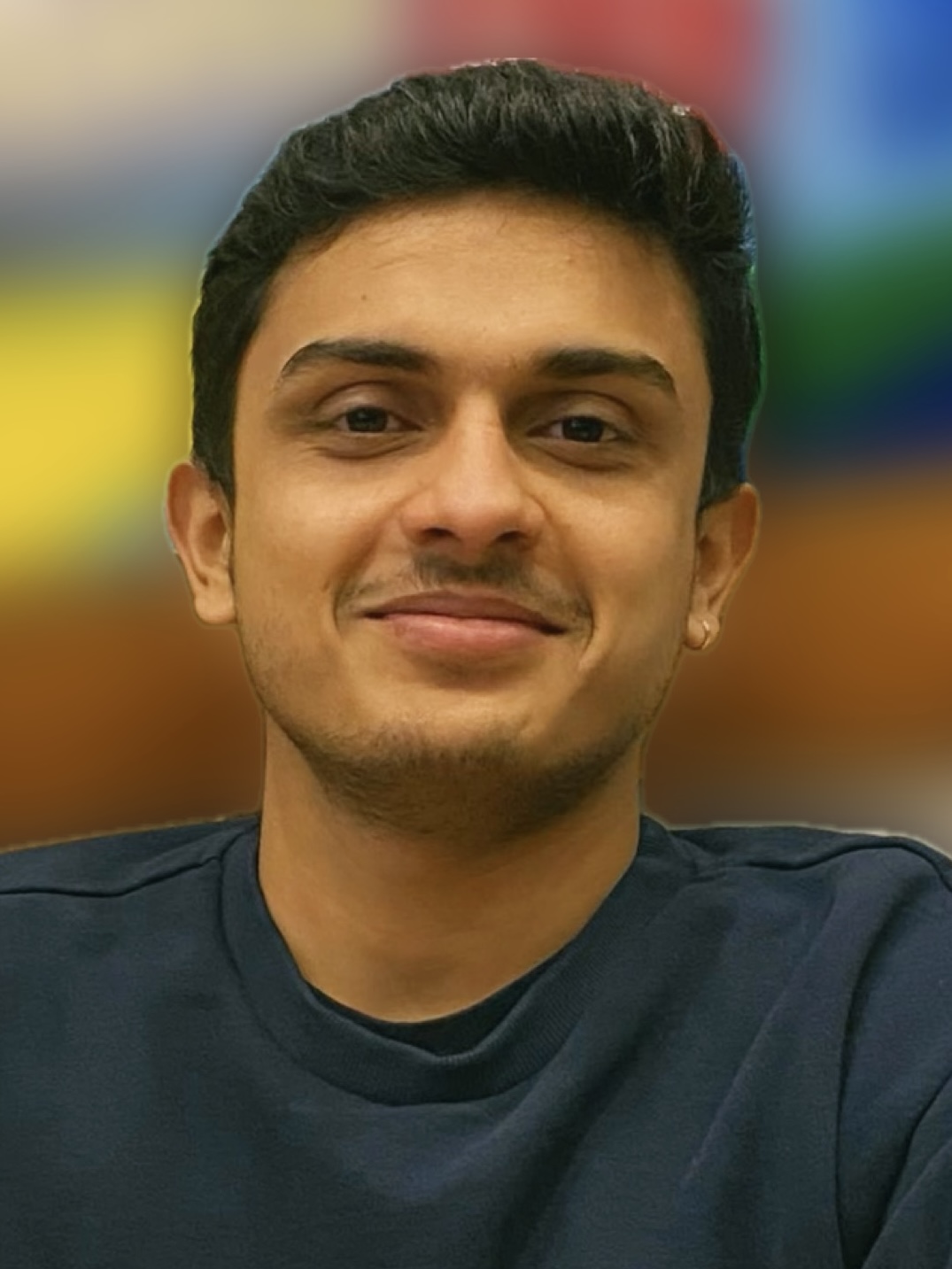}}]{Sathwik Chadaga}
(Student Member, IEEE) received the dual degree (B.Tech. and M.Tech.) in electrical engineering from Indian Institute of Technology Madras, Chennai, India, in 2020, and the M.S. degree in aeronautics and astronautics from the Massachusetts Institute of Technology, Cambridge, MA, USA, in 2023, where he is currently pursuing the Ph.D. degree with the Laboratory for Information and Decision Systems. His current research interests include stochastic optimization and online learning in control of network systems. 
\end{IEEEbiography}
\begin{IEEEbiography}[{\includegraphics[width=1in,height=1.25in,clip,keepaspectratio]{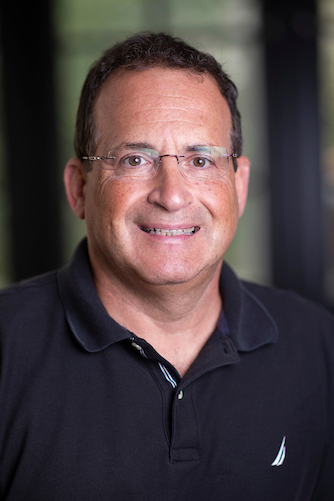}}]{Eytan Modiano}
	is The Richard C. Maclaurin Professor in the Department of Aeronautics and Astronautics and the Laboratory for Information and Decision Systems (LIDS) at MIT.  Prior to Joining the faculty at MIT in 1999, he was a Naval Research Laboratory Fellow between 1987 and 1992, a National Research Council Post Doctoral Fellow during 1992-1993, and a member of the technical staff at  MIT Lincoln Laboratory between 1993 and 1999.  Eytan Modiano received his B.S. degree in Electrical Engineering and Computer Science from the University of Connecticut at Storrs in 1986 and his M.S. and PhD degrees, both in Electrical Engineering, from the University of Maryland, College Park, MD, in 1989 and 1992 respectively.

	His research is on modeling, analysis and design of communication networks and protocols.    He received the Infocom Achievement Award (2020) for contributions to the analysis and design of cross-layer resource allocation algorithms for wireless, optical, and satellite networks.   He is the co-recipient of the Infocom 2018 Best paper award, the MobiHoc 2018 best paper award, the MobiHoc 2016 best paper award, the Wiopt 2013 best paper award, and the Sigmetrics 2006 best paper award.  He was the Editor-in-Chief for IEEE/ACM Transactions on Networking (2017-2020), and served as Associate Editor for IEEE Transactions on Information Theory and IEEE/ACM Transactions on Networking.  He was the Technical Program co-chair for  IEEE Wiopt 2006, IEEE Infocom 2007, ACM MobiHoc 2007, and DRCN 2015; and general co-chair of Wiopt 2021.  He had served on the IEEE Fellows committee in 2014 and 2015, and is a Fellow of the IEEE and an Associate Fellow of the AIAA.
\end{IEEEbiography}
\vfill

\end{document}